\documentclass[reqno]{amsproc}

\usepackage{amstext,amsmath,amssymb,amsfonts}
\usepackage[latin1]{inputenc}
\usepackage{epsfig}
\usepackage{hyperref}
\usepackage{color}

\usepackage{latexsym}

\newcommand{\bea}{\begin{eqnarray}}
\newcommand{\eea}{\end{eqnarray}}
\newcommand{\beq}{\begin{equation}}
\newcommand{\eeq}{\end{equation}}
\newcommand{\enn}{\nonumber \end{equation}}

 \newcommand{\cG}{\mathcal{G}}

  \newcommand{\cJ}{\mathcal{J}}

\newcommand{\cZ}{{\mathcal{Z}}}
\newcommand{\cW}{{\mathcal{W}}}

\newcommand{\cH}{{\mathcal H}}

\newcommand{\tr}{{\rm Tr}}

\title[Renormalization and Hopf Algebraic Structure for the quartic melonic model ]{Renormalization and Hopf Algebraic Structure\\ of the  5-Dimensional Quartic Tensor Field Theory}

\author{Remi C. Avohou}
\address[R.C.A.]{
International Chair in Mathematical Physics and Applications,
ICMPA-UNESCO Chair, 072BP50, Cotonou, Rep. of Benin}
\email{avohouremicocou@yahoo.fr}

\author{Vincent Rivasseau}
\address[V.R.]{
Laboratoire de Physique Th\'eorique, CNRS UMR 8627, Universit\'e Paris 11, Paris-Saclay, 91405 Orsay Cedex, France and Perimeter Institute for Theoretical Physics, 31 Caroline St. N, N2L 2Y5, Waterloo, ON, Canada.}
\email{rivass@th.u-psud.fr}

\author{Adrian Tanasa}
\address[A.T.]{LIPN, CNRS UMR 7030,
Universit\'e Paris 13, Sorbonne Paris Cit\'e,  93430 Villetaneuse, France and Horia Hulubei National Institute for Physics and Nuclear Engineering P.O.B. MG-6 077125 Magurele, Romania.}
\email{adrian.tanasa@ens-lyon.org}

\begin{document}

\maketitle 
\begin{abstract} This paper is devoted to the study of renormalization 
of the quartic melonic tensor model in dimension (=rank) five. We review the perturbative renormalization and the computation of the 
one loop beta function, confirming the asymptotic freedom of the model. We then define the Connes-Kreimer-like Hopf algebra 
describing the combinatorics of the renormalization of this model
and we analyze in detail, at one- and two-loop levels, the Hochschild cohomology allowing to write the combinatorial Dyson-Schwinger equations. Feynman tensor graph Hopf subalgebras are also exhibited.
\end{abstract}

\tableofcontents
\section{Introduction}

Random tensor models have in the recent years witnessed a major revival.  Tensor models studied in the 90's \cite{tens1,tens2,tens3} had only
$U(N)$ symmetry under \emph{simultaneous identical} change of basis on every tensor index.
The consideration of more general, non-symmetric ``colored" tensors \cite{Gurau:2009tw,Gurau:2011xp} and of related 
``uncolored models" \cite{Gurau:2011kk,Bonzom:2012hw} (see also \cite{ABH} for models interpolating between the colored and uncolored cases)
lead to extend this symmetry into a $U(N)^{\otimes D}$ symmetry 
under \emph{independent changes on each of the tensor indices}. 
It allowed to discover an associated tensorial $1/N$ expansion \cite{Gurau:2010ba}. 
This expansion is governed by a new integer, the Gurau degree. It is \emph{not} a topological invariant of the dual triangulated manifold, but 
the sum of the genera of a canonical set of Heegaard surfaces called \emph{jackets}. The basic combinatorial invariants
which are both the vertices and observable of rank $d$ models and the Feynman graphs of rank $d-1$ models are the bipartite
$d$-regular edge colored graphs. Among the initial results of the theory are the existence of single \cite{Bonzom:2011zz} 
and double \cite{Dartois:2013sra,GurauSchaeffer,Bonzom:2014oua} scaling limits. 

In dimension/rank three, the closely related multi-orientable model \cite{Tanasa:2011ur} can sum over a larger class of graphs (see also \cite{praa} for a short review). It also admits a $1/N$
expansion with single  and double scaling limits \cite{Dartois:2013he,dubla}.

Tensor models perform sums over triangulations in arbitrary dimension pondered by a discretization of
the Einstein-Hilbert action, hence they can be considered the equilateral version of Regge calculus \cite{ambjorn}. 
The \emph{tensor track} \cite{Rivasseau:2013uca} therefore 
proposes such models as candidates for the quantization of gravity in dimensions higher than two. In particular
when slightly broken at the propagator level, the $1/N$ tensorial expansion provides power counting rules for an associated class of non-local 
quantum field theories, which generalize non-commutative field theories, in particular the Grosse-Wulkenhaar model \cite{Grosse:2004yu} and the translation-invariant model \cite{GMRT}. 
Renormalizable models have been defined in this class \cite{BGR,Geloun:2013saa}, including a family of models \cite{Carrozza:2013wda,Lahoche:2015ola}
which incorporate the gauge (=trivial holonomy) constraints of group field theory \cite{Oriti:2011jm}. 
Asymptotic freedom is a generic property of such models, at least when only quartic interactions are present \cite{BenGeloun:2012pu}, in contrast with
asymptotic safety for non-commutative (=matrix) models of the Grosse-Wulkenhaar type \cite{Disertori:2006nq}. 
The situation is more complicated for models with higher order interactions \cite{BenGeloun:2012yk,Carrozza:2014rba}.
Recently the numerical exploration of renormalization flows in the tensor \emph{theory space} 
\cite{Rivasseau:2014ima}  has started, using the Wetterich equation \cite{Benedetti:2014qsa}, 
in order to discover interesting new random geometries.

Among other popular mathematical tools to analyze renormalization are the Connes-Kreimer Hopf algebra \cite{CK} 
and its Hochschild cohomology \cite{anatomy}.
The Connes-Kreimer structures describes in a purely algebraic way the combinatorics of renormalization.
Moreover, appropriate Hochschild one-cocyles allow to encode, in a recursive way, the combinatorics of the Dyson-Schwinger equation - the so-called combinatorial Dyson-Schwinger equation (see \cite{teza-yeats} and references within).
 Connes-Kreimer-like Hopf algebras have been defined for some renormalizable Moyal \cite{fab}, \cite{TK} and sixth order interaction tensor models \cite{matti}. 
 The combinatorial Dyson-Schwinger equation has also been analyzed for Moyal QFT in \cite{TK}. For the sake of completeness, 
 let us also mention that, using different perspectives, studies of the Dyson-Schwinger in the tensor model setting have been performed in \cite{Bonzom:2014oua} and \cite{thomas}.

\medskip

In this paper we deepen the analysis of one of the simplest renormalizable tensor field theory, namely the 
dimension/rank five tensor model with usual propagator $1/(p^2 + m^2)$ and quartic melonic interactions. It is just renormalizable,
and has been studied in the melonic approximation, including some numerical analysis, in \cite{Samary:2014oya}.
We give the power counting of general graphs, recall the self-consistency equations of the melonic sector \cite{Samary:2014oya}, and check that except in the vacuum sector only melonic graphs diverge. We also check 
the asymptotic freedom of the model through a detailed one-loop computation. We then turn on to define the Connes-Kreimer algebra for this model and the Hochschild one-cocycles allowing to write the combinatorial Dyson-Schwinger equations. We show the corresponding diagrams at one- and two-loop levels. Furthermore, Hopf subalgebras are exhibited, again at one- and two-loop levels.

\section{The Model}

\subsection{The Bare Model}

Consider a pair of conjugate rank-5 tensor fields 
\bea T_n, \bar T _{\bar n},\;\;  {\rm  with} \;\; n = \{ n_1,n_2,n_3,n_4,n_5 \} \in \mathbb{Z}^5, 
\;\;  \bar n = \{\bar n_1,\bar n_2,\bar n_3,\bar n_4 , \bar n_5 \} \in \mathbb{Z}^5.
\eea 
They belong respectively to the tensor product 
$\cH = \cH_1 \otimes \cH_2 \otimes\cH_3\otimes\cH_4\otimes\cH_5$ and to its dual,
where each $\cH_c$ is an independent copy of  $\ell_2 (\mathbb{Z})= L_2 (U(1))$, and  the color or strand index $c$ takes values $c=1,2,3,4,5$.
By Fourier transform the field $T$ can be considered also as an ordinary complex scalar field  
$T  (\theta_{1},\theta_{2},\theta_{3}, \theta_4, \theta_5)$ on the five-dimensional torus ${\rm \bf T}_5 = U(1)^5$
and $\bar T(\bar \theta_{1},\bar \theta_{2},\bar \theta_{3}, \bar \theta_4, \bar \theta_5 )$ is simply its complex conjugate 
 \cite{BGR}. The tensor indices $n$ can be identified as the \emph{momenta} associated to 
the positions $\theta$, and we therefore also call \emph{momentum basis} the canonical basis $e_n= e_{n_1}  \otimes \cdots \otimes e_{n_5} $ of $\cH$.

If we restrict the momenta $n$ to lie in $[-N, N]^5$ rather than in $\mathbb{Z}^5$ we have a proper (finite dimensional) tensor model.
We consider $N$ as the ultraviolet cutoff, and we are interested in performing the ultraviolet limit $N \to \infty$.

\begin{figure}[h]
 \centering
     \begin{minipage}[t]{.8\textwidth}
      \centering
\includegraphics[angle=0, width=4cm, height=1cm]{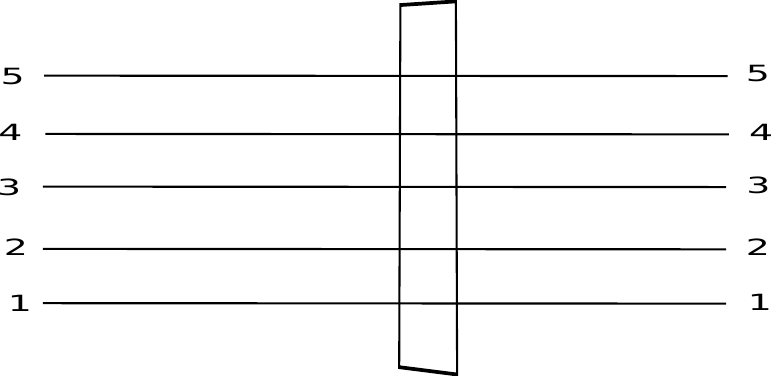}
\vspace{0.1cm}
\caption{ {\small The strands represent the five different colors and the box suggests the $C$ propagator that links them.}}
\label{fig:propa}
\end{minipage}
\put(-110,12){$\bar T$}
\put(-240,12){$T$}
\end{figure}

We introduce the normalized Gaussian measure
\bea
d\mu_{C}(T, \bar T) = \left(\prod_{n, \bar n \in [-N, N]^5} \frac{dT_n d\bar T_{\bar n}}{2i\pi} \right) \mathrm{Det}( C )^{-1} \ 
e^{-\sum_{n} T_n C^{-1} (n) \bar T_{n}},
\eea
where the bare covariance $C$ is, up to a bare field strength parameter $Z$, the inverse of the Laplacian on ${\rm \bf T}_5$ 
with cutoff $N$ on each momentum plus a bare mass term
\bea
C (n,\bar n )= \delta_{n,\bar n} C (n) , \quad C (n) = \frac{1}{Z} \frac{1}{n^2+ m_b^2}.
\eea
In such formulas letters such as $n,m,p,q, \cdots$ will be used for elements in $[-N, N]^5$,
$n^2 = \sum_c n_c^2$ and $m_b^2$ is the square of the bare mass. Figure \ref{fig:propa} gives a graphical representation of  the propagator of this theory.

We adopt the color notations of \cite{Delepouve:2014bma}. $\tr$, $ \mathbb{I}$ and $<,>$ mean the trace, the identity and the scalar product on 
$\cH$. $ \mathbb{I}_{c}$ is the identity on $\cH_c$, $\tr_c$ is the trace on $\cH_c$
and $<,>_{c}$ the scalar product restricted to $\cH_{c}$. 
The notation $\hat c$ means `every color except $c$". For instance 
$\cH_{\hat c}$ means $\otimes_{c' \ne c}  \cH_{c'}$, $ \mathbb{I}_{\hat c}$ is the identity on the tensor product $\cH_{\hat c}$,
$\tr_{\hat c}$ is the partial trace over $\cH_{\hat c}$ and $<,>_{\hat c}$ the scalar product restricted to $\cH_{\hat c}$.

$T$ and $\bar T$ can be considered both as vectors in $\cH$ or as diagonal 
(in the momentum basis) operators acting on $\cH$, with eigenvalues $T_n$ and $\bar T_n$. An important quantity in melonic tensor models
is the partial trace $\tr_{\hat c}  T \bar T$, which we can also identify with the partial product $<T , \bar T  >_{\hat c} $. It is
a (in general non-diagonal) operator in $\cH_c$ with matrix elements in the momentum basis
\bea <T , \bar T  >_{\hat c} (n_c, \bar n_c)  = ( \tr_{\hat c}  T \bar T )(n_c, \bar n_c) 
= \sum_{n_{c'}, \bar n_{c'}, c'\ne c} \prod_{c'\neq c}\delta_{n_{c'} \bar n_{c'}}T_{n} \bar T_{\bar n}.
\eea

\begin{figure}[h]
 \centering
     \begin{minipage}[t]{.8\textwidth}
      \centering
\includegraphics[angle=0, width=3cm, height=2cm]{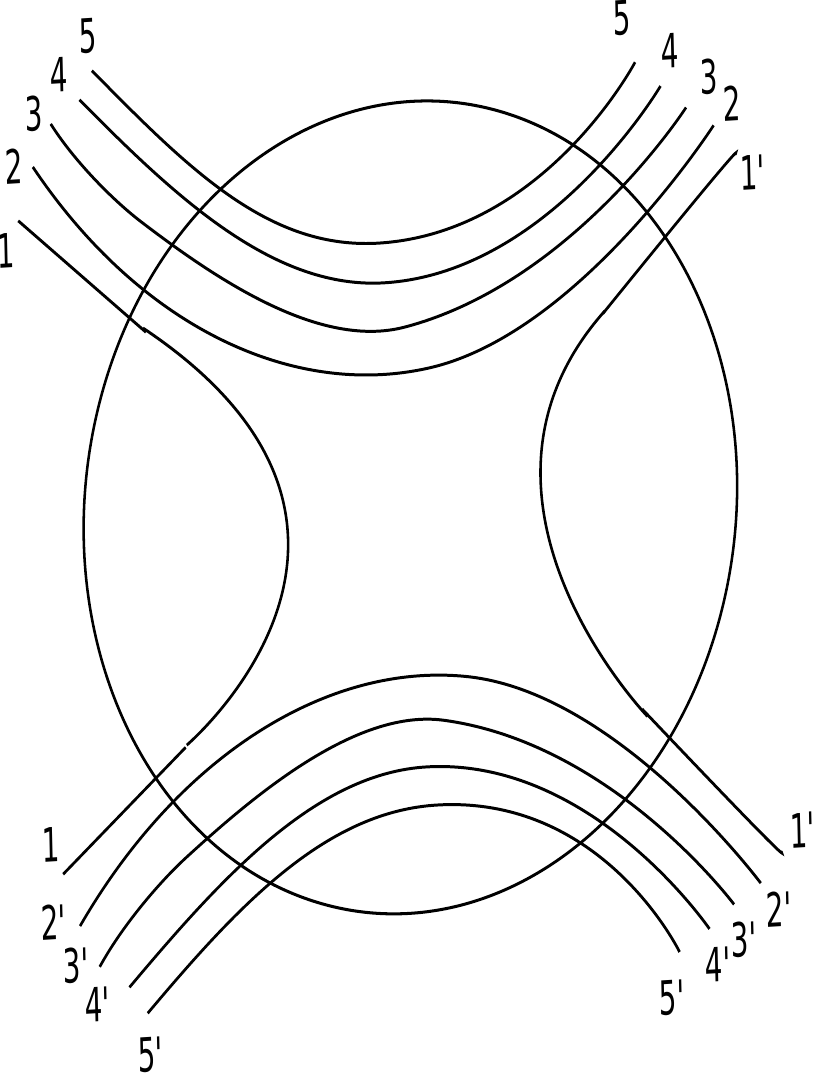}\hskip1cm\includegraphics[angle=0, width=3cm, height=2cm]{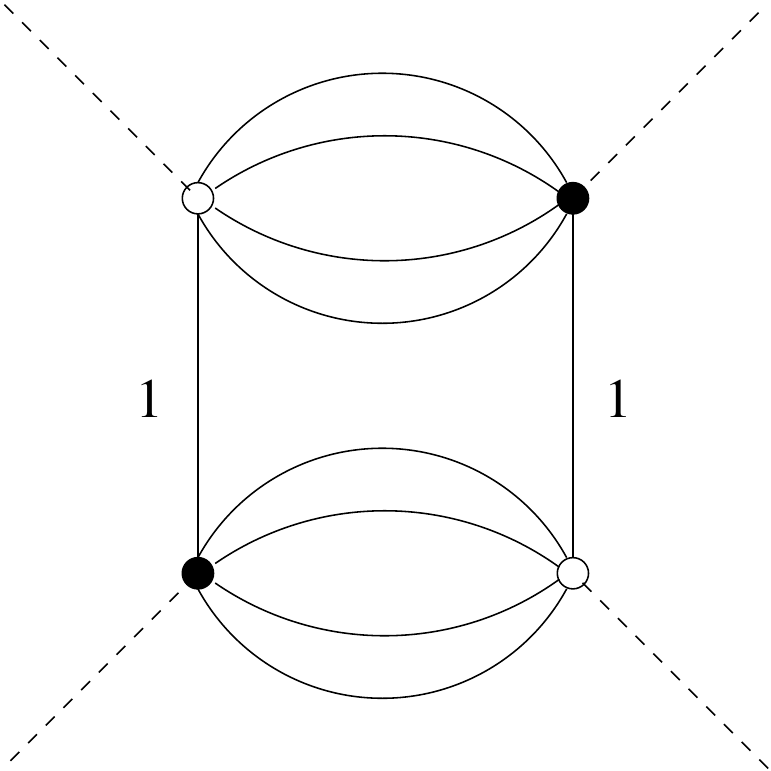}
\vspace{0.1cm}
\caption{ {\small Two equivalent representations of the melonic quartic vertex of type $V_{1}$}.}
\label{fig:vertex}
\end{minipage}
\end{figure}

The main new feature of tensor models compared to ordinary field theories 
is the non-local form of their interaction, which is chosen invariant under the action of $U(N)^{\otimes 5}$.
In this paper we consider only the quartic melonic interaction, which is 
$\frac{g_b}{2} \sum_c V_c(T, \bar T) $ where $g_b$ is the bare coupling constant and
\bea
V_c(T, \bar T) = \tr_c  [ ( \tr_{\hat c}  T \bar T )^2 ] =
\sum_{n_c,\bar n_c, m_c,\bar m_c}  \left( T_n\bar T_{\bar n}  \prod_{c'\neq c}\delta_{n_{c'} \bar n_{c'}} \right) \delta_{n_c \bar m_c} \delta_{m_c \bar n_c} 
\left(\ T_m\bar T_{\bar m} \prod_{c'\neq c}\delta_{m_{c'} \bar m_{c'}}   \right)  \label{formulavc}
\eea  
for $c= 1, \cdots , 5$ are the five quartic melonic interactions of random tensors at rank five \cite{Delepouve:2014bma}  (see Figure \ref{fig:vertex}).

The generating function for the moments of the model is then
\bea
\cZ(g,J, \bar J)=  \frac{1}{\cZ}\int e^{<\bar J , T> + <J , \bar T >}   e^{-g_b \sum_c V_c(T, \bar T) } d\mu_{C}(T, \bar T) ,
\eea
where $\cZ= \cZ(g,J, \bar J)\vert_{J =\bar J =0}$ is the normalization and the source tensors $J$ and $\bar J$
are dual respectively to $\bar T$ and $T$. The main problem in quantum field theory is to compute
$\cW(g,J, \bar J)  = \log  \cZ(g,J, \bar J)$ which is the generating function for the connected Schwinger functions
\bea  S_{2k} (\bar n_1 , \cdots , \bar n_k; n_1 , \cdots,  n_k ) = \frac{\partial}{\partial J_{\bar n_1}} \cdots \frac{\partial}{\partial J_{\bar n_k}}  
 \frac{\partial}{\partial \bar J_{n_1}} \cdots \frac{\partial}{\partial J_{n_k}}  \cW(g,J, \bar J) \vert_{J= \bar J =0} .
\eea
This model is globally symmetric under color permutations and is just renormalizable like ordinary $\phi^4_4$. 
However the structure of ultraviolet divergencies is simpler, as shown in the next subsection. 

\subsection{Divergent graphs}
A way to determine the divergent degree $\omega(\cG)=-2L(\cG)+F(\cG)$ of a graph $\cG$, where $L(\cG)$ and $F(\cG)$ are respectively the
 number edges and faces of $\cG$, is by computing the number of its faces.
This quantity is given by 
\bea
F=4V + 4-2E -\frac{1}{12}(\sum_\cJ g_{\tilde \cJ}-\sum_{\cJ_\partial}g_{\cJ_\partial})-(C_{\partial\cG}-1), \label{jacketsetc}
\eea
where $V$ and $E$  are respectively the number of vertices and the number of external legs  of $\cG$. 
The sum is performed on all jackets $\cJ$ of $\cG_{color}$, $g_{\tilde \cJ}$ is the genus of the pinched jacket $\tilde \cJ$, $g_{\cJ_\partial}$ 
is the genus of the jacket $\cJ_\partial$ and $C_{\partial\cG}$ is the number of connected components of the boundary graph $\partial\cG$.
For precise definitions of all these notions and a proof of \eqref{jacketsetc}, we refer the interested reader to \cite{BGR,matti,Samary:2014oya}.

After substituting the combinatorial relation $2L+E=4V_4$, the divergent degree of $\cG$ can be written as
\bea
\omega(\cG)= 4-E - (C_{\partial\cG}-1) - \frac{1}{12}(\sum_\cJ g_{\tilde \cJ}-\sum_{\cJ_\partial}g_{\cJ_\partial}).
\eea

We call $1$PI graphs $\cG$ with $\omega(\cG)\geq 0$ {\it superficially divergent}, and write $\cG_{sd}^{\omega} := \{\cG : \omega(\cG)\geq 0\}$ for the set of superficially divergent $1PI$
Feynman graphs.

In the following table, we give a list of superficially divergent graphs, all over cases are finite as $N \to \infty$.

\begin{center}
\begin{tabular}{lcccc||cc|}
\label{tabel}
$E$ & $\sum_{J_\partial}g_{J_\partial}$ & $C_{\partial\cG}-1$ & $\sum_{\tilde J}  g_{\tilde J}$ & $\omega(\cG)$  \\
\hline\hline
4 & 0 & 0 & 0 & 0 \\
\hline
2 & 0 & 0 & 0 & 2\\
\hline
0 & 0 & -1 & 0 & 5\\
\hline
0 & 0 & -1 & 36 & 2\\
\hline
0 & 0 & -1 & 60 & 0\\
\hline
\hline
\end{tabular} 
\medskip

Table 1: List of superficially divergent graphs
\end{center}

\section{Renormalization and $\beta$ function}

The theory is perturbatively just renormalizable and asymptotically free \cite{Samary:2014oya}.
However the structure of divergent subgraphs is simpler both than in ordinary $\phi^4_4$ or in the Grosse-Wulkenhaar model and its translation-invariant renormalizable version.  

\begin{figure}[!h]
\begin{center}
{\includegraphics[height=2cm]{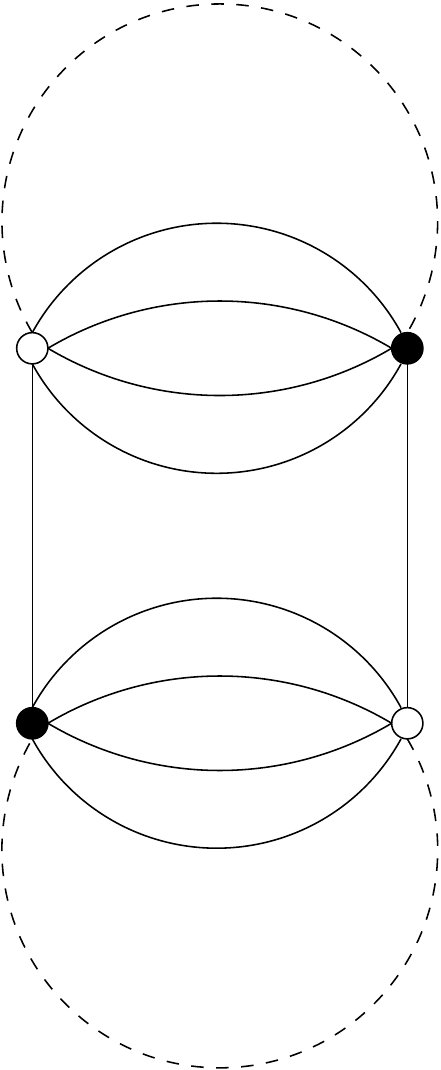}} \hskip.3cm {\includegraphics[height=2cm]{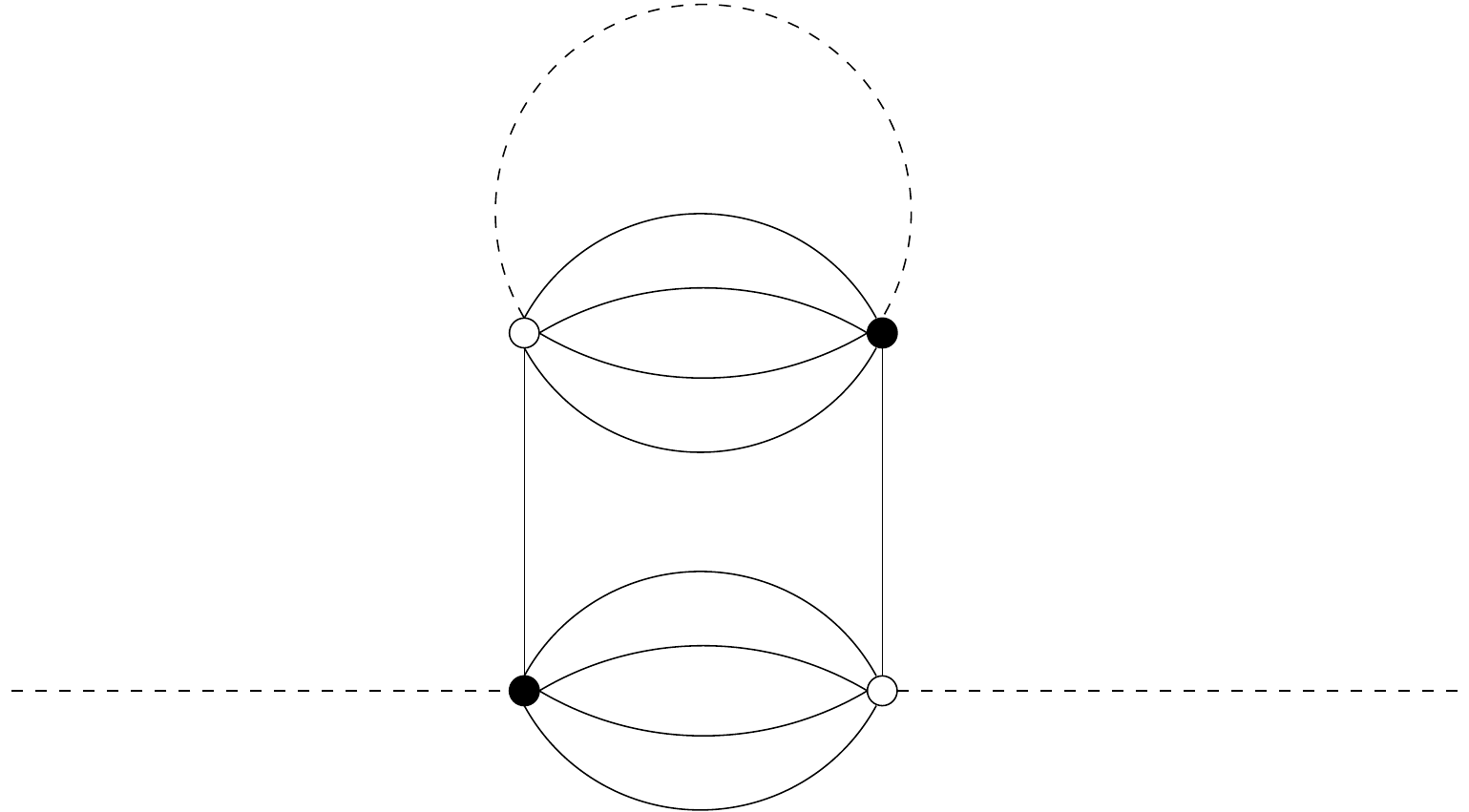}}\hskip.5cm
{\includegraphics[height=1.5cm,angle=0]{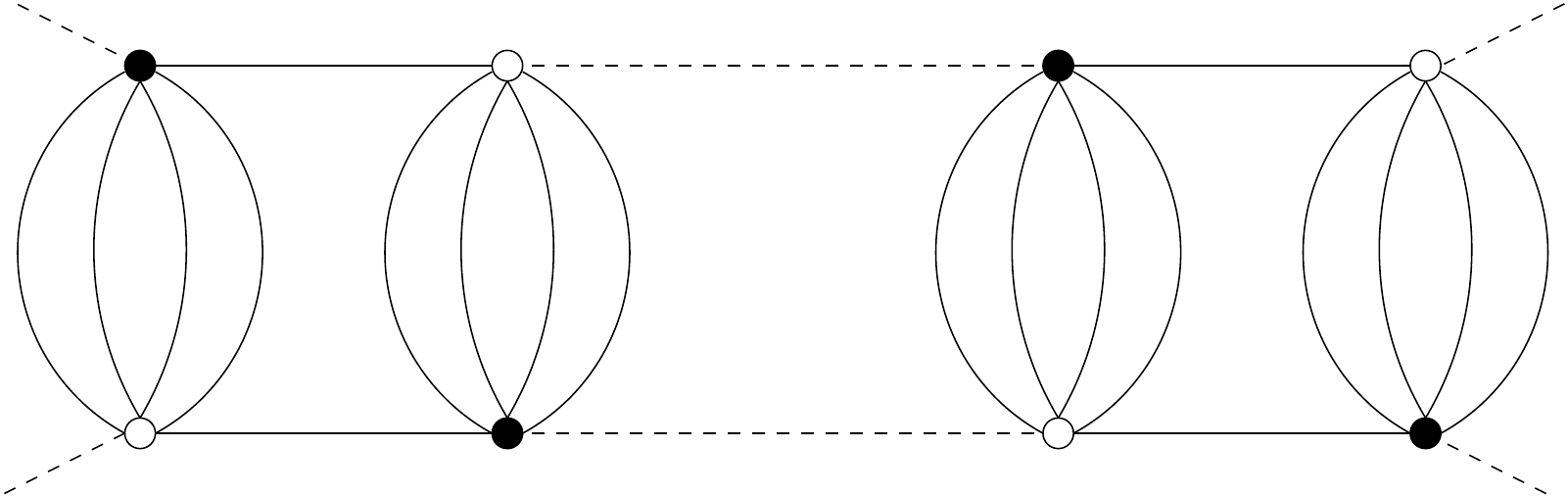}}
\end{center}
\caption{From left to right, the fundamental melons for the 0, 2 and 4 point function.
The melonic quartic vertex is shown with plain edges, and the dashed edges correspond to Wick contractions of $T$ with $\bar T $,
hence bear an inverse Laplacian.}
\label{melonicdivergences}
\end{figure}

\begin{figure}[!h]
\begin{center}
{\includegraphics[height=1.7cm,angle=0]{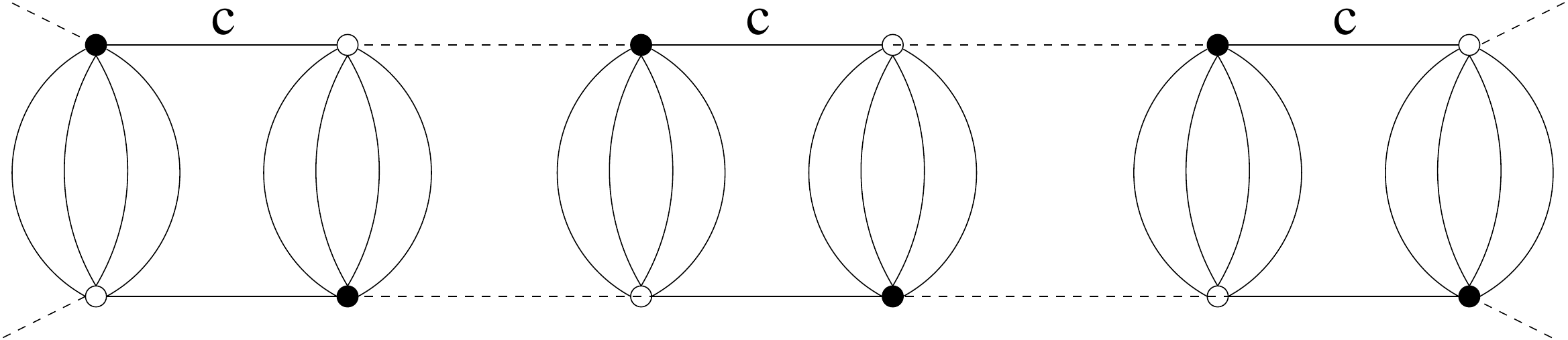}}
\end{center}
\caption{The length-two melonic four-point chain. Remark that for the graph to diverge,
all vertices along the chain must be of same color type.}
\label{fourpointmelonicdivergences}
\end{figure}

Melonic graphs with 0, 2 and 4 external legs are divergent, respectively as $N^5$, $N^2$ and $\log N$.
They are obtained respectively from the fundamental melonic graphs of Figure \ref{melonicdivergences}, 
by recursively inserting the fundamental 2-point melon on any dotted line, or, in the case of the four point function, also replacing any vertex by the fundamental
4-point melon so as to create a ``melonic chain" of arbitrary length (see Figure \ref{fourpointmelonicdivergences} for a chain of length 2), in which all vertices must be of the same color (otherwise the graph won't be divergent).

\begin{figure}[!h]
\begin{center}
{\includegraphics[height=2.3cm,angle=0]{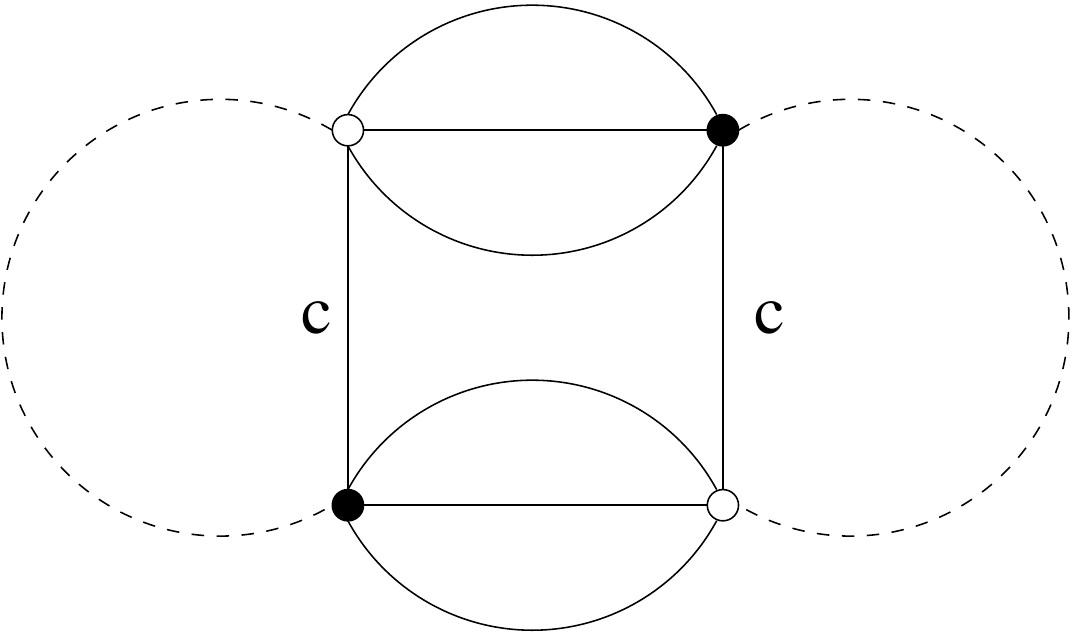}}\hskip2cm{\includegraphics[height=2.6cm]{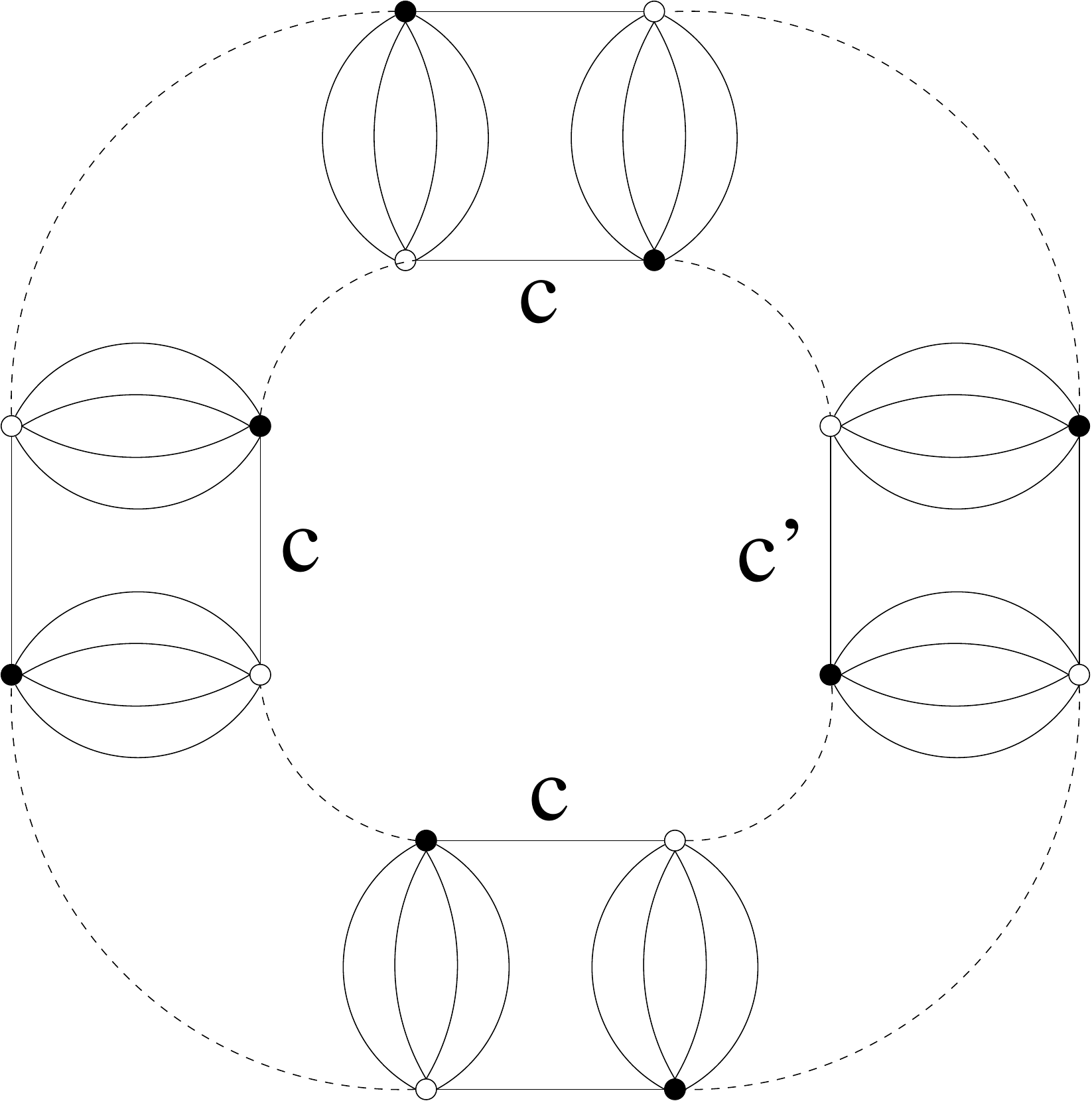}}
\end{center}
\caption{A length-one and a length-four non-melonic divergent vacuum connected necklace. Remark that the left necklace diverges a $N^2$,
whereas  the right one diverges as $\log N$ for $c \ne c'$.}
\label{vacuumnonmelonicdivergences}
\end{figure}

Beyond melonic approximation there is only one simple infinite family of non melonic graphs who are divergent. They are vacuum graphs divergent  either as $N^2$ or as $\log N$.
They are made of a "necklace chain" of arbitrary length $p\ge 1$, decorated with arbitrary
melonic insertions. Two such necklace chains, of length 1 and 4, are pictured in Figure \ref{vacuumnonmelonicdivergences}. If all couplings along the chains have same color
the divergence is quadratic, in $N^2$. If some couplings are different, the divergence is logarithmic, in $\log N$.

In the rest of this paper we shall no longer further consider the vacuum graphs, as they are not physically observable.

\subsection{Melonic Sector}

Let us call $G_E^{mel}$ and $\boldsymbol\Gamma_E^{mel}$ respectively the connected and one-particle irreducible melonic functions (i.e. sum over the melonic
Feynman amplitudes) of the theory  with $E$ external fields. The
bare melonic two point function $G_{2,b}^{mel} (n, \bar n ) = \delta(n, \bar n)G_{2,b}^{mel} (n)$ is related to the bare melonic self-energy $\boldsymbol\Gamma_{2,b}^{mel} (n, \bar n) = 
\delta(n, \bar n) \boldsymbol\Gamma_{2,b}^{mel} (n)$ by the usual equation
\bea
G_{2,b}^{mel} (n) = C_b(n) \frac{1}{1 -  C_b(n) \boldsymbol\Gamma_{2,b}^{mel} (n)}     \label{g2}.
\eea
$\boldsymbol\Gamma_{2,b}^{mel} (n)$ is the sum over colors $c$ of a unique
function $\Gamma_{2,b}^{mel}$ of the single integer $n_c$:
\bea \boldsymbol \Gamma_{2,b}^{mel} (n) = \sum_c\Gamma_{2,b}^{mel} (n_c). \label{sigma2}
\eea
$\boldsymbol\Gamma^{mel}_{2,b}$ is uniquely defined by \eqref{g2}-\eqref{sigma2} and the closed equation
\bea
 \Gamma_{2,b}^{mel} (n_c ) = - 2g_b  \sum_{p } \delta (p_c -n_c)  G_{2,b}^{mel} (p). \label{gamma2}
\eea
Indeed the combinatoric coefficient for the single loop two-point melonic graph in the center of Figure \ref{melonicdivergences} is 1 since
it has a single vertex with weight $-g_b$, and two Wick contractions, as we have to choose 
which of the two melonic pairs of the single vertex is contracted together. This leads to the factor $-2g_b$ in \eqref{gamma2}.

Similarly the \emph{bare} melonic four point vertex function $\boldsymbol\Gamma_{4,b}^{mel} (n,\bar n, m, \bar m)$ is the sum over colors $c$ of contributions
defined through a unique matrix $ \Gamma_{4,b}^{mel} (n_c, m_c)$ which corresponds to the melonic invariant $V_c$:
\bea  \boldsymbol\Gamma_{4,b}^{mel} (n,\bar n, m, \bar m) = - g \sum_c    \delta (n_{\hat c}  , \bar n_{\hat c}  ) 
\delta (m_{\hat c}  , \bar m_{\hat c}  )   \delta (n_{c}  , \bar m_{c}  )  \delta (m_{c}  , \bar n_{c}  ) 
 \Gamma_{4,b}^{mel} (n_c,m_c). \label{sigma4}
\eea
$\boldsymbol\Gamma^{mel}_{4,b}$ is uniquely defined by \eqref{sigma4} and the closed equation
\bea
\Gamma_{4,b}^{mel} (n_c ,m_c) = 1  - 2g_b  \sum_{p, q }\delta (p_c -n_c)  G_{2,b}^{mel} (p)  \delta (q_c -m_c)  G_{2,b}^{mel} (q)  \Gamma_{4,b}^{mel} (n_c ,m_c),\label{gamma4}
\eea
which solves to
\bea
 \Gamma_{4,b}^{mel} (n_c ,m_c) = \frac{1}{1 + 2g_b  \sum_{p, q }\delta (p_c -n_c)  G_{2,b}^{mel} (p)  \delta (q_c -m_c)  G_{2,b}^{mel} (q) }.\label{gamma4sol}
\eea
Indeed this closed equation expresses the sum over a chain of arbitrary length generalizing the single loop melonic four point graph in the right of Figure \ref{melonicdivergences},
which is the solution of expanding the denominator in \eqref{gamma4sol} as a geometric series. 
This chain is decorated with ``cactus" melonic insertions on each dotted propagator, which come
from expanding the $G_{2,b}^{mel}$ functions in \eqref{gamma4sol} according to \eqref{g2}\footnote{This prescription can be described in even simpler terms in the intermediate field representation 
(see eg \cite{Lahoche:2015ola}) where melonic graphs become simple trees.}.
It remains to check the overall combinatoric coefficient 1 in front of $g_b$ in \eqref{gamma4}-\eqref{gamma4sol}. 
This coefficient can be checked by the following reasoning: adding a vertex at the end of a chain of length $n$ creates
a chain of length $n+1$; the new vertex has two melonic pairs and a factor $-g_b$, and there are again two Wick contractions 
corresponding to the choice of the melonic pair $(T, \bar T)$ which contracts to the chain\footnote{See \cite{Gurau:2013pca} for a more detailed 
study of the combinatoric coefficients of melonic functions with arbitrarily many external legs in this type of quartic model.}. 

At fixed cutoff $N$ these equations define $\boldsymbol\Gamma^{mel}_{2,b}$, $G_{2,b}^{mel} $ and $\boldsymbol\Gamma^{mel}_{4,b}$ (hence also $G_{4,b}^{mel}$)
at least as analytic functions for $g_b$ sufficiently small, 
because the number of melonic graphs is exponentially bounded as the number of vertices increases. However this does not allow to take the 
limit as $N \to \infty$ since the radius of convergence shrinks to zero in this limit: we need to now renormalize.

\subsection{Perturbative Renormalization}

The renormalization is given in terms of a melonic-BPHZ scheme which is given by BPHZ-like normalization conditions at zero external momenta, 
but restricted to the divergent sector, namely melonic graphs\footnote{The BPHZ prescription in standard renormalizable QFT imposes conditions on the full 1PI functions of the theory, 
not just their melonic part. This is because all 1PI graphs diverge in standard field theory. In this tensorial theory since non-melonic graphs are convergent the full 
BPHZ prescription is not minimal, and differs from the melonic BPHZ prescription only by unnecessary finite renormalizations.}.

The standard renormalization procedure expresses $\Gamma_{2,b}^{mel}$ in terms of
renormalized quantities through a Taylor expansion 
\bea
\label{one}
\Gamma_{2,b}^{mel} ( n)=\Gamma_{2,b}^{mel} ( 0)+  n^2\frac{\partial \Gamma_{2,b}^{mel} }{\partial n^2} \Big|_{n=0}+
\Gamma_{2,b}^{mel,r} (n) =(Z-1) n^2+Zm_b^2-  m_{r}^2+\Gamma_{2,b}^{mel,r} ( n),
\eea
with
\begin{eqnarray}\label{eq:conz}
m_b^2=\frac{m_{r}^2+\Gamma_{2,b}^{mel} (0) }{Z},\quad Z=1+\frac{\partial \Gamma_{2,b}^{mel} }{\partial n^2} \Big|_{n=0}.
\end{eqnarray}
We have the usual renormalization conditions on the self-energy \cite{Samary:2014oya}:
\beq\label{con}
\Gamma_{2,b}^{mel,r} (0)=0,\quad \frac{\partial \Gamma_{2,b}^{mel,r}}{\partial n^2}\Big|_{n=0}=0 .
\eeq
$Z$ is the (bare) wave function normalization, which can be exchanged for rescaling the bare fields as $T \to Z^{-1/2} T$, $\bar T \to Z^{-1/2} T$. 
The melonic two-point function is therefore 
\bea G_{2,b}^{mel}(n)  =  \frac{1}{Z (n^2 + m_b^2)  -  (Z-1) n^2 - Zm_b^2  +  m_{r}^2 - \Gamma_{2,b}^{mel,r} ( n)}  =  \frac{1}{n^2 + m_{r}^2 -\Gamma_{2,b}^{mel,r} ( n)}.\label{eq:gamma2}
\eea
The renormalized coupling $g_r$ is then defined through the usual renormalization condition at zero momenta applied to any of the five terms of the color sum 
for $\boldsymbol\Gamma_{4,b}^{mel} (0,0,0,0)$ in \eqref{sigma4}.   Hence
\begin{equation}
g_r =g_b\Gamma_{4,b}^{mel} (0,0).\label{gren}
\end{equation}

\subsection{One Loop $\beta$ function}

In quartic models of this type the \emph{bare} wave-function renormalization $Z$ can be absorbed into rescaling bare fields as $T \to Z^{-1/2} T$, $\bar T \to Z^{-1/2} T$ so that the acton
in terms of the rescaled fields has rescaled bare coupling $g_b \to Z^{-2} g_b$. At fixed renormalized coupling, we would like to identify
how this rescaled bare coupling flows with the cutoff $N$. The relationship at one loop is
\bea Z^{-2} g_b = g_r  + g^2_r [\beta_2 \log N + {\rm finite} ]   + O(g_r^3), \label{renflow}
\eea
where $\beta_2$ is the one-loop coefficient of the beta function. As well-known $\beta_2 >0$ corresponds to an exploding flow in the ultraviolet (``Landau ghost") but 
$\beta_2 <0$ corresponds to the nice physical situation of \emph{asymptotic freedom}: the (rescaled) bare coupling 
flows to zero as $N \to \infty$, so that the (rescaled) theory becomes closer and closer to a Gaussian (free) QFT in the ultraviolet regime. 
Remembering \eqref{gren}, we can also reexpress \eqref{renflow} as 
\bea 
Z^2\Gamma_{4,b}^{mel} (0 ,0) =  1 -  g_b [\beta_2 \log N + {\rm finite} ]   + O(g_b^2)  .\label{beta0}
\eea
Hence to compute $\beta_2$ we can expand $\Gamma_{4,b}^{mel} (0 ,0)$ and  $Z-1$ at first order, neglecting the $O(g_b^2)$ contributions. 
Using the equations \eqref{sigma4}, \eqref{gamma4sol} and \eqref{eq:gamma2}, we have 
\bea
\Gamma_{4,b}^{mel} (0 ,0)   = 1 -  2g_b \sum_{p\in[-N,N]^4}\frac{1}{(p^2+m_r^2)^2}+O(g_b^2) .
\label{beta1}
\eea

From the second equation in \eqref{eq:conz} joined to the equations \eqref{sigma2}, \eqref{gamma2} and \eqref{eq:gamma2} we have
\bea
Z =1 + \frac{\partial \Gamma_{2,b}^{mel} }{\partial n^2} \Big|_{n=0}= 1  +2 g_b \sum_{p\in[-N,N]^4}\frac{1}{(p^2+m_r^2)^2}+O(g_b^2).  \label{beta2}
\eea
Indeed there are two minus signs which compensate, one in front of $g_b$ in \eqref{gamma2} and the other coming from the mass soustraction (or equivalently, derivation of a denominator):
\bea
\frac{1}{n_c^2}   \bigl( \frac{1}{p^2 + n_c^2  + m_r^2}  - \frac{1}{p^2  + m_r^2}  \bigr)  \vert_{n_c = 0} = - \frac{1}{ ( p^2 + m_r^2 )^2} .
\eea

The following identity computes the exact coefficient of the logarithmic divergence in $N$ of the momentum sum  $\sum_{p\in[-N,N]^4}\frac{1}{(p^2+m_r^2)^2}$.
\bea
\sum_{p\in[-N,N]^4}\frac{1}{(p^2+m_r^2)^2} &= 2 \pi^2 \log N + f(N)   . \label{beta3}
\eea
with $f(N)$ bounded as $N \to \infty$.
This identity is proved in the Appendix.

Combining \eqref{beta1}-\eqref{beta3} we get
\bea
Z^2 \Gamma_{4,b}^{mel} (0,0)  =   1 +  2g_b [2\pi^2 \log N + f(N)]   + O(g_b^2) .
\eea

Hence comparing with \eqref{beta0}, with this normalization convention
\bea 
\beta_{one\; loop} =\beta_{2} = - 4\pi^2 <0 ,
\eea
which means that the theory is asymptotically free, as expected for such a quartic melonic renormalizable model, in agreement with  
\cite{BenGeloun:2012pu,Benedetti:2014qsa,Samary:2014oya,Lahoche:2015ola}.

\section{Algebraic properties of the combinatorics of renormalization}
\subsection{Hopf algebra}

Let us first recall, from \cite{matti}, the notion of a residue of a tensor graph. A {\bf residue} $Res(\cG)$  of a tensor 
$1$PI Feynman graph of our model is obtained by contracting all internal edges of $\cG$ to a point. 

 Let $\mathcal{H}$ be the vector space freely generated by the $1$PI Feynman tensor graphs of our model (including the empty graph, which we denote by $1_{\mathcal{H}}$).
We endow this vector space with a product given by the disjoint union 
and with a unit map $\eta:\mathbb{K}\to\mathcal{H}$ defined by $\eta(1) = 1_{\mathcal{H}}$ and zero elsewhere, where we have denoted by $\mathbb{K}$ a field such as $\mathbb{R}$ or $\mathbb{C}$.
We further endow $\mathcal{H}$
with the coproduct $\Delta: \mathcal{H}\rightarrow\mathcal{H}\otimes\mathcal{H}$ 
given by
\bea\label{eq:coproduct}
\Delta(\cG) := \cG\otimes 1_{\mathcal{H}} + 1_{\mathcal{H}}\otimes \cG +\sum_{\gamma}\gamma\otimes\cG/\gamma,
\eea
where the sum is taken on disjoint unions of superficially divergent subgraphs 
of $\cG$.

Furthermore, we define $\Delta'$ to be the non-trivial part of the coproduct:
$$\Delta(\cG)=\cG\otimes 1_{\mathcal{H}} + 1_{\mathcal{H}}\otimes\cG + \Delta'(\cG).$$
Note that $\mathcal{H}$ is graded by the loop number. One can then prove that the coproduct above is coassociative. This follows \cite{matti} - one needs to check that the set of $1$PI Feynman graphs of the model (see Table $1$) 
is closed under tensor subgraph contraction and insertion.

The counit $\epsilon:\mathcal{H}\to\mathbb{K}$ is defined by: $\epsilon(1_{\mathcal{H}}) = 1$ and $\epsilon(\cG) = 0$ for all $\cG\neq 1_{\mathcal{H}}$.
Defining now the antipode in the usual way  one concludes that $(\mathcal{H}, \Delta)$ is a Hopf algebra, which is actually the Connes-Kreimer Hopf algebra describing the combinatorics of the renormalization of 
our five-dimensional quartic tensor models
(see again \cite{matti} for details).

\subsection{Combinatorial Dyson-Schwinger equation and Hochschild cohomology}

In this subsection, we show that a primitively divergent graph $\gamma$ can be used to define a Hochschild $1-$cocycle $B_+^{\gamma}$, where  the operator $B_+^{\gamma}$ is given by the operation of insertion of tensor graphs into $\gamma$. We then use these operators to encode, in a recursive way, the combinatorics of the Dyson-Schwinger equation.

We denote by $|\cG|$ the number of loops of $\cG$.
We then define the operators $B_+:\mathcal{H}\to \mathcal{H}$ by
\bea
\label{defbplus}
B_+^{\gamma}(X) :=  \sum_{\cG}\frac{bij(\gamma,X,\cG)}{|X|_V}\frac{1}{maxf(\cG)}\frac{1}{(\gamma|X)}\cG,\quad
B_+^{k;r} = \sum_{|\gamma|=k,\, res(\gamma)=r}B_+^{\gamma},
\eea
where $maxf(\cG)$ is the number of ways to shrink subdivergencies such that the resulting graph is primitive and $|X|_V$ is the number of graphs which are equal upon removal of external edges. The quantity $bij(\gamma,X,\cG)$ is the number of bijections between the external edges of $X$ and adjacent edges of places in $\gamma$ such that $\cG$ is obtained. Finally, the notation $\gamma|X$ stands for the number of insertion places of the graph $X$ in $\gamma$.

We further set
\bea
B_+^{\gamma}(1_{\mathcal{H}}) = \gamma,
\eea
and
\bea
\label{defc}
h_k^r := \sum_{|\cG|=k, res(\cG) = r}\cG
\eea
where the sum runs on the graphs of a given loop number and residue. Finally, let $M_r$ be the set of graphs such that $res(\cG)=r$.

In commutative or Moyal non-commutative scalar QFT, 
(see \cite{anatomy} or resp. \cite{TK}) for any Feynman graph $X_{k,r}$ with $k$ loops and  residue $r$, one has the following identities:
\bea
\label{unu}
\cG^r\equiv1+\sum_{\cG\in M_r}\cG = 1 + \sum_{k=1}^\infty 
B_{+}^{k,r}(X_{k,r}),
\eea
\bea
\label{doi}
 \Delta(B_+^{k;r}(X_{k,r})) = B_+^{k;r}(X_{k,r})\otimes \mathbb{I}+ (id_{\mathcal{H}}\otimes B_+^{k;r})\Delta (X_{k,r}),
\eea
\bea
\label{trei}
 \Delta(h_k^r)=\sum_{j=0}^k Pol_j^r(c)\otimes h_{k-j}^r,
 \eea 
 where $Pol_j^r(c)$ is a polynomial in the variables $h_m^r$ of total degree $j$.

Let us now mention the meaning of each of these three identities. The first one encodes, in a recursive way, the combinatorics of the Dyson-Schwinger equation - the so-called combinatorial Dyson-Schwinger equation. The second one points out the fact that the operator $B_+$ is, from a purely algebraic point of view, a Hochschild cocycle of the Hopf algebra of the previous subsection. Finally, the last equation shows that one can define Hopf subalgebras at any loop order.

\medskip

In the rest of this subsection, we will explicitly check under what conditions these algebraic identities hold at the one- and two-loop level, in the case of our five-dimensional quartic tensor model. We set for simplicity, $\includegraphics[angle=0, width=1.5cm, height=0.8cm]{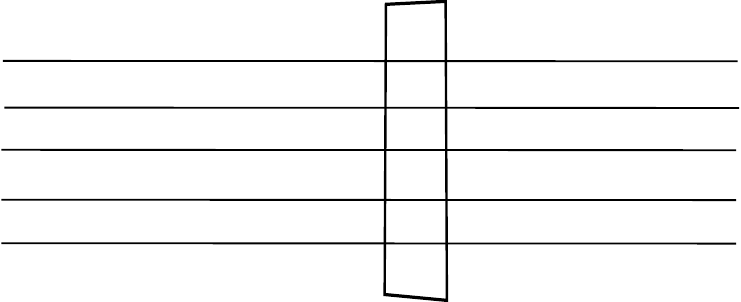}=\includegraphics[angle=0, width=1cm, height=0.1cm]{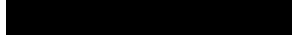}$ and $\includegraphics[angle=0, width=1.5cm, height=1.4cm]{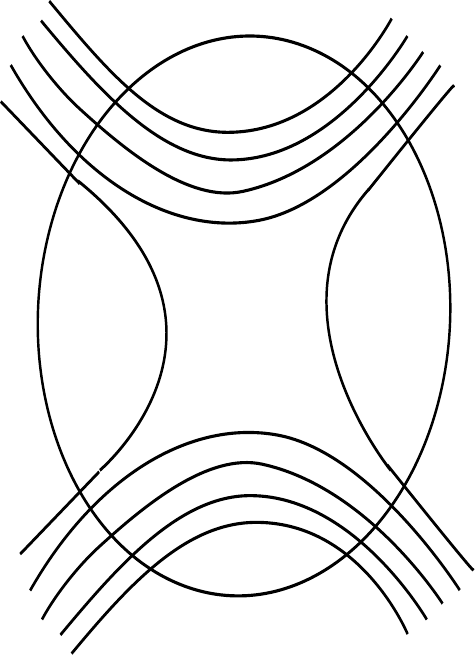}=\includegraphics[angle=0, width=1.5cm, height=1.4cm]{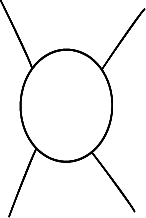}$.

$\bullet$ {\bf One-loop analysis:}
We have
\bea
B_+^{1,\includegraphics[angle=0, width=1cm, height=0.1cm]{propagatorgras.pdf}}=\sum_cB_+^{\includegraphics[angle=0, width=1.5cm, height=1.4cm]{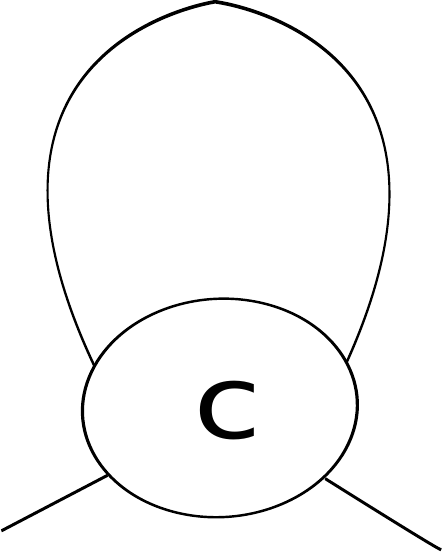}} \quad \mbox{ and } B_+^{1,\includegraphics[angle=0, width=1.5cm, height=1.4cm]{simplevertex.pdf}}=\sum_cB_+^{\includegraphics[angle=0, width=1.5cm, height=1.4cm]{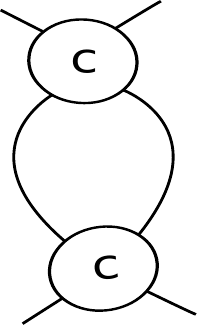}},
\eea
where the sum $\sum_c$ runs over all five colors, and by $\includegraphics[angle=0, width=1.5cm, height=1.4cm]{oneloopc.pdf}$ one means $\includegraphics[angle=0, width=1.5cm, height=1.4cm]{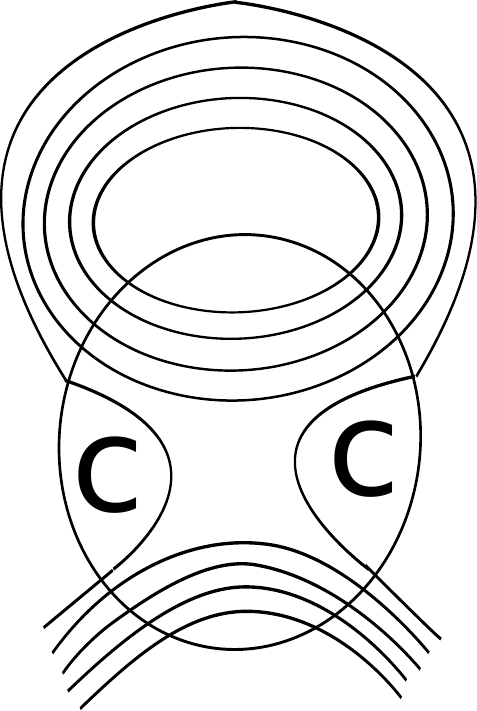}$ the 
quartic vertex $V_c$ with special thread of color $c$ (see Figure \ref{fig:vertex} for the case $c=1$). 
Respectively by $\includegraphics[angle=0, width=1.5cm, height=1.4cm]{graphh3.pdf}$ one means $\includegraphics[angle=0, width=1.5cm, height=1.4cm]{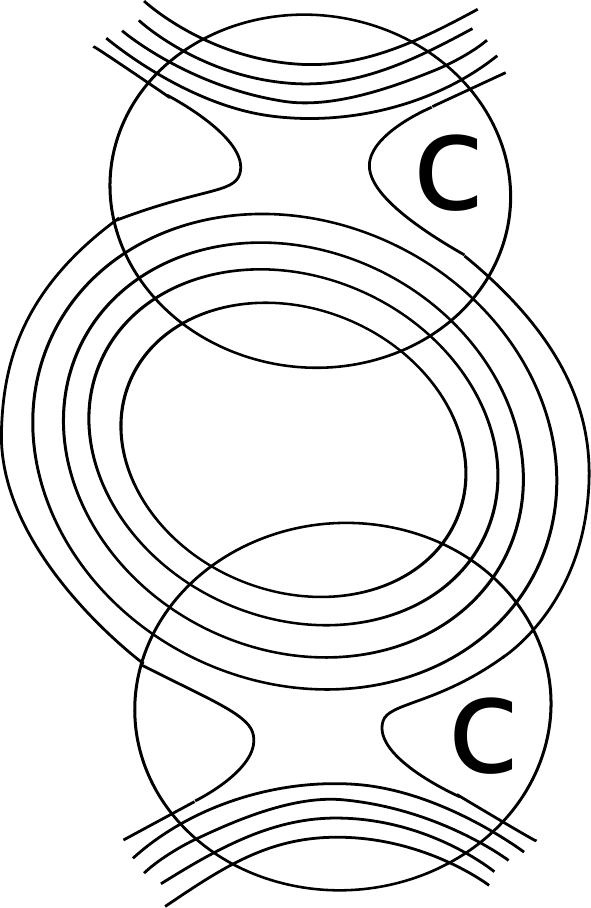}$.

One then has:
\bea
h_1^{\includegraphics[angle=0, width=1cm, height=0.1cm]{propagatorgras.pdf}}=B_+^{1,\includegraphics[angle=0, width=1cm, height=0.1cm]{propagatorgras.pdf}}(1_{\mathcal{H}}) = \sum_c
\includegraphics[angle=0, width=1.5cm, height=1.4cm]{oneloopc.pdf}
\eea
and
\bea \label{eq:c1}
h_1^{\includegraphics[angle=0, width=1.5cm, height=1.4cm]{simplevertex.pdf}}=B_+^{1,\includegraphics[angle=0, width=1.5cm, height=1.4cm]{simplevertex.pdf}}(1_{\mathcal{H}})=\sum_c\includegraphics[angle=0, width=1.5cm, height=1.4cm]{graphh3.pdf}.
\eea

$\bullet$ {\bf Two-loop analysis:}
As above, one computes
\bea\label{eq:42}
h_2^{\includegraphics[angle=0, width=1cm, height=0.1cm]{propagatorgras.pdf}}=\sum_{c,c'}\includegraphics[angle=0, width=1.5cm, height=1.4cm]{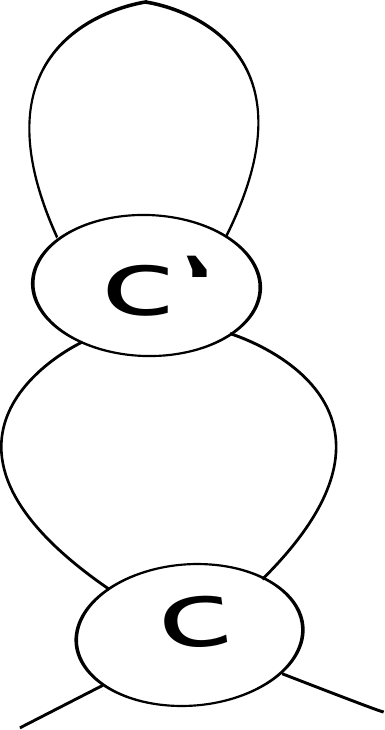},
\eea
where the sum over $c$ and $c'$ of $\includegraphics[angle=0, width=1.5cm, height=1.4cm]{graphh4.pdf}$ now corresponds to two different types of stranded diagrams. If $c=c'$ we get
$\includegraphics[angle=0, width=1.5cm, height=1.4cm]{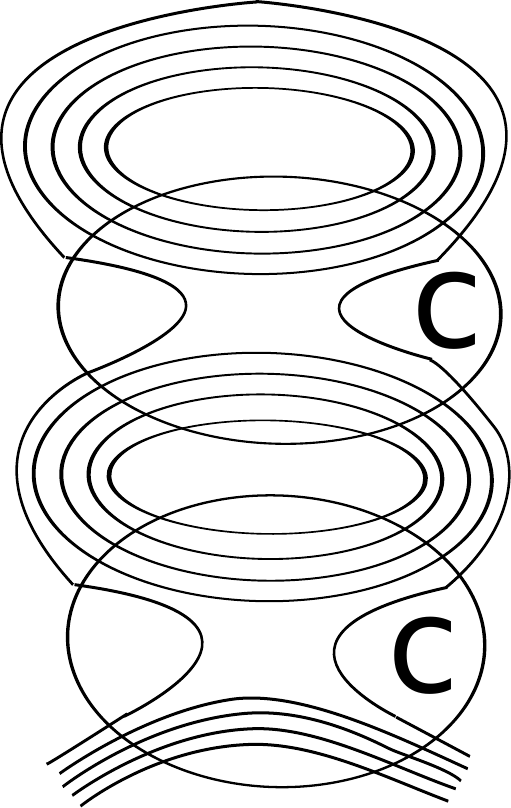}$. If $c\neq c'$, we get a different
diagram namely $\includegraphics[angle=0, width=1.5cm, height=1.4cm]{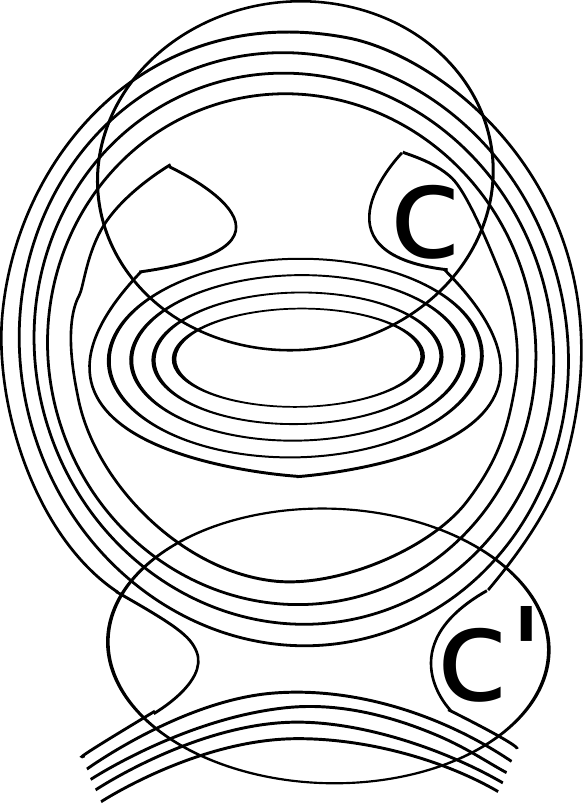}$. Both types of graphs are quadratically divergent, since they have both eight inner closed faces and three propagators.
Remark however that they do not have the same amplitude: the coefficients in front of the quadratic divergence are different. 

The relation in \eqref{eq:42} coincides with the diagrammatic expression obtained directly from definition \eqref{defc}.
Let us now apply $\Delta$ on each of the graphs in the RHS of \eqref{eq:42}. For each of these tensor graphs, the non-trivial part of the coproduct writes: 
\bea\label{eq:43}
\Delta'(\includegraphics[angle=0, width=1.5cm, height=1.4cm]{graphh4.pdf})=\includegraphics[angle=0, width=1.5cm, height=1.4cm]{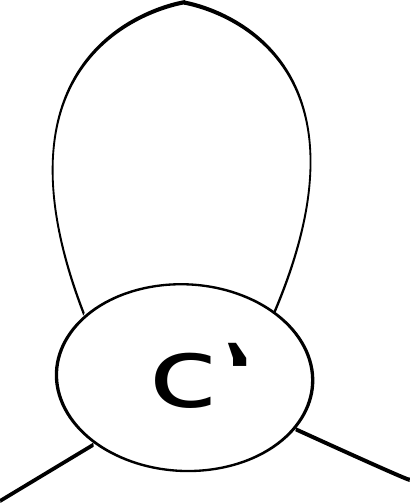}\otimes \includegraphics[angle=0, width=1.5cm, height=1.4cm]{oneloopc.pdf},
\eea
when ($c\neq c'$) and 
\bea\label{eq:43}
\Delta'(\includegraphics[angle=0, width=1.5cm, height=1.4cm]{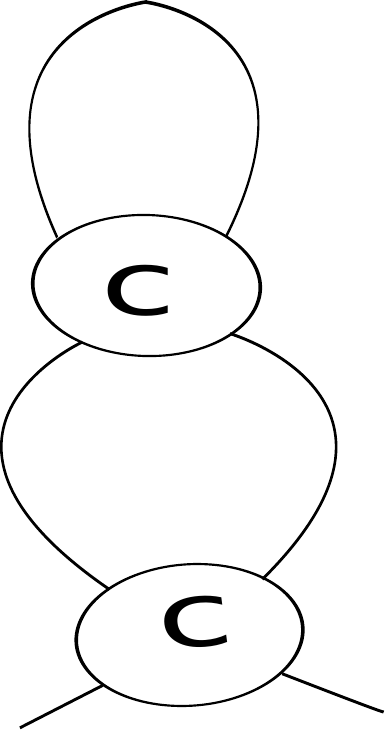})=\includegraphics[angle=0, width=1.5cm, height=1.4cm]{oneloopc.pdf}\otimes \includegraphics[angle=0, width=1.5cm, height=1.4cm]{oneloopc.pdf} + \includegraphics[angle=0, width=1.5cm, height=1.4cm]{graphh3.pdf}\otimes \includegraphics[angle=0, width=1.5cm, height=1.4cm]{oneloopc.pdf},
\eea
else. As example, we have:
\bea\label{eq:43}
\Delta'(\includegraphics[angle=0, width=1.5cm, height=1.4cm]{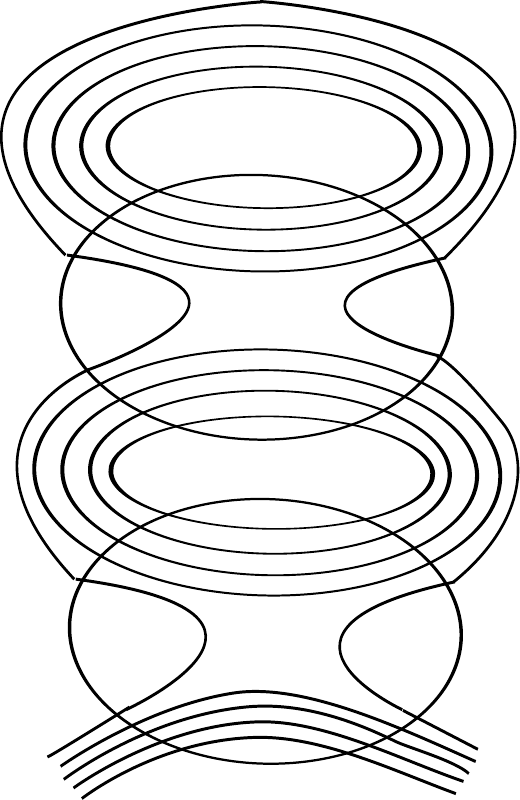})=\includegraphics[angle=0, width=1.5cm, height=1.4cm]{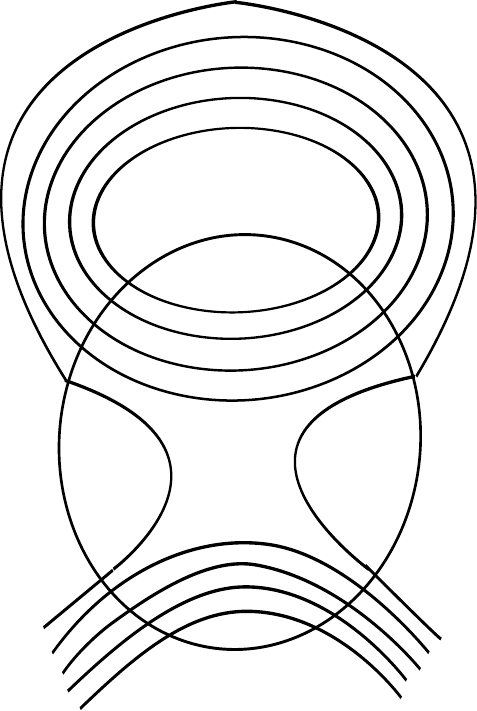}\otimes \includegraphics[angle=0, width=1.5cm, height=1.4cm]{graph94.pdf}+\includegraphics[angle=0, width=1.5cm, height=1.4cm]{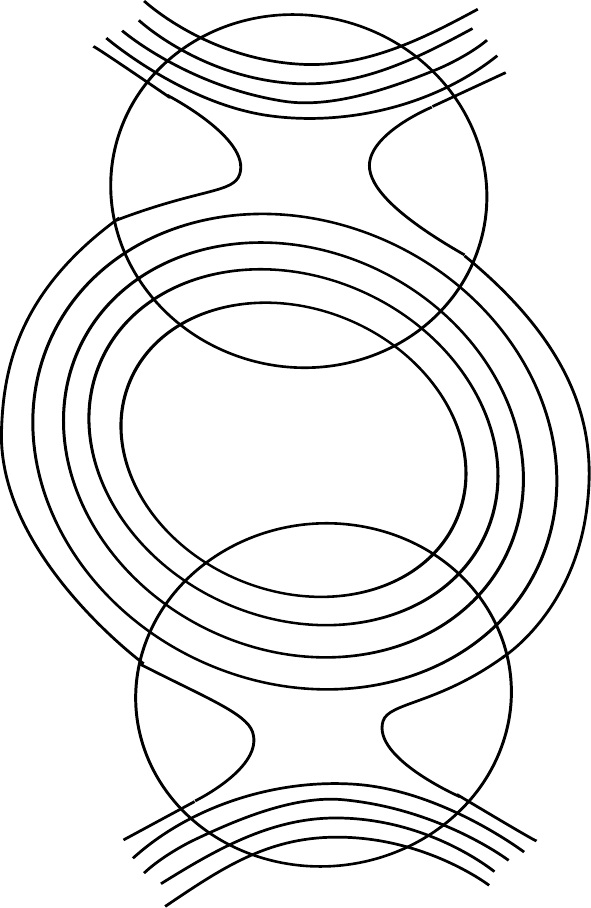}\otimes \includegraphics[angle=0, width=1.5cm, height=1.4cm]{graph94.pdf}.
\eea
Adding up all these contributions, we obtain
\bea\label{eq:relationc1}
\Delta'(h_2^{\includegraphics[angle=0, width=1cm, height=0.1cm]{propagatorgras.pdf}})=(h_1^{\includegraphics[angle=0, width=1cm, height=0.1cm]{propagatorgras.pdf}}+h_1^{\includegraphics[angle=001, width=1.5cm, height=1.4cm]{simplevertex.pdf}})\otimes h_1^{\includegraphics[angle=0, width=1cm, height=0.1cm]{propagatorgras.pdf}},
\eea
which thus checks relation \eqref{trei} at two-loop level.

\medskip

For the four-point function, we have
\bea\label{4p}
h_2^{\includegraphics[angle=0, width=1.5cm, height=1.4cm]{simplevertex.pdf}}=\sum_{c}\includegraphics[angle=0, width=1.5cm, height=1.6cm]{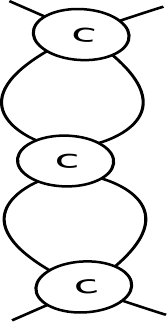} + 2\sum_{c,c'} \includegraphics[angle=0, width=1.5cm, height=1.4cm]{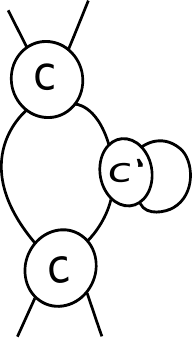},
\eea
where  $\includegraphics[angle=0, width=1.5cm, height=1.6cm]{graphh6.pdf}$ corresponds to $\includegraphics[angle=0, width=1.5cm, height=1.4cm]{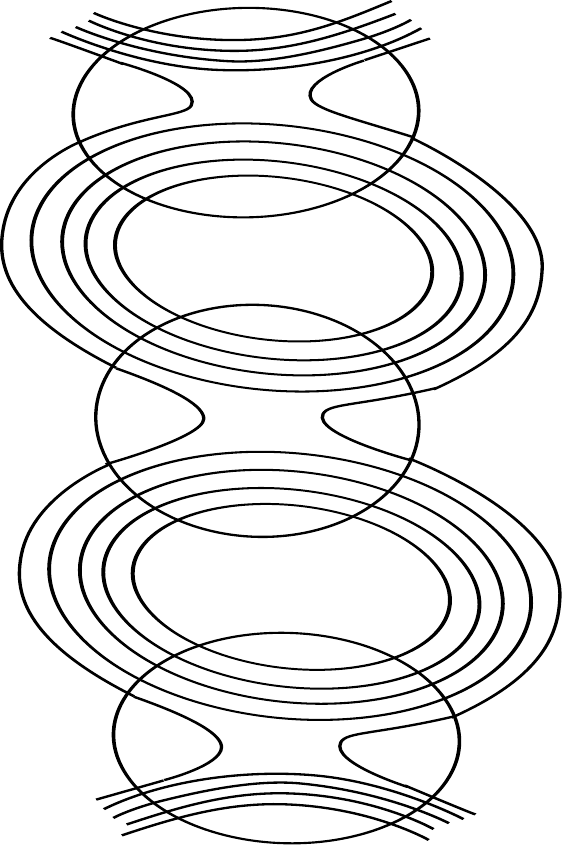}$, and
by $\includegraphics[angle=0, width=1.5cm, height=1.4cm]{graphh7.pdf}$ one means $\includegraphics[angle=0, width=1.5cm, height=1.4cm]{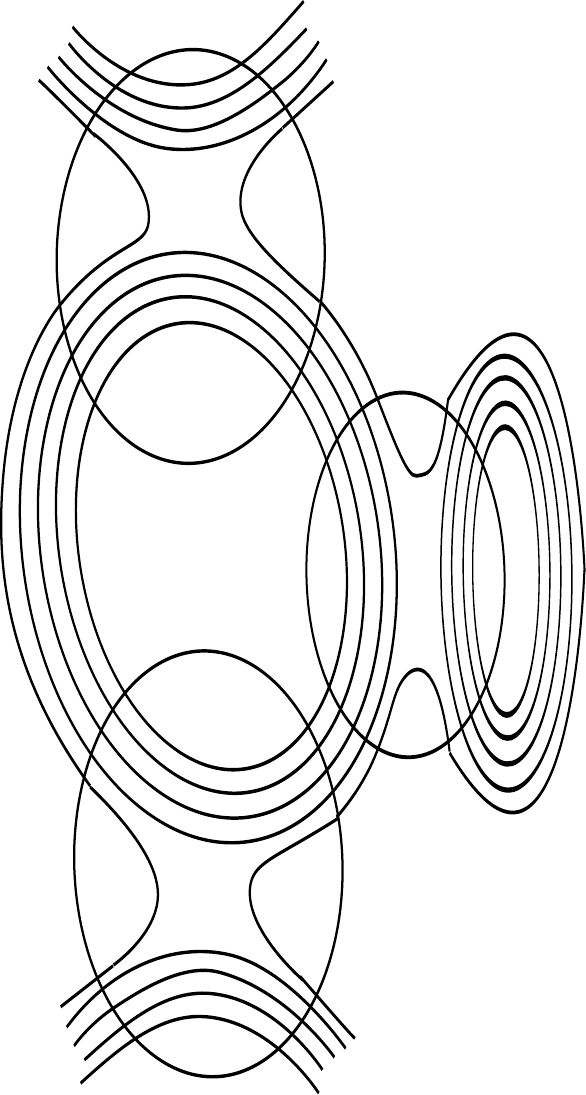}$ (for $c=c'$) or $\includegraphics[angle=0, width=1.5cm, height=1.4cm]{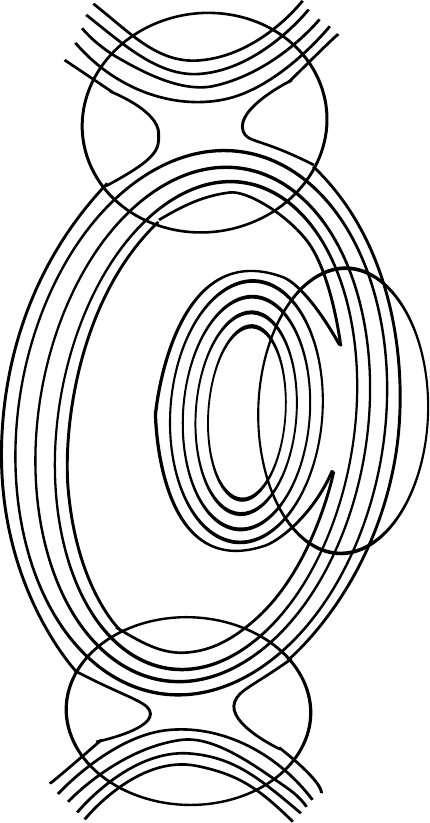}$ for $c \ne c'$.

Once again, we apply the coproduct on each of these tensor graphs and we obtain for the non-trivial part
\bea\label{eq:deltac21}
\Delta'(\includegraphics[angle=0, width=1.5cm, height=1.4cm]{graph101.pdf})=2\includegraphics[angle=0, width=1.5cm, height=1.4cm]{graph96.pdf}\otimes \includegraphics[angle=0, width=1.5cm, height=1.4cm]{graph96.pdf}, \quad \Delta'(\includegraphics[angle=0, width=1.5cm, height=1.4cm]{graph102.pdf})=\includegraphics[angle=0, width=1.5cm, height=1.4cm]{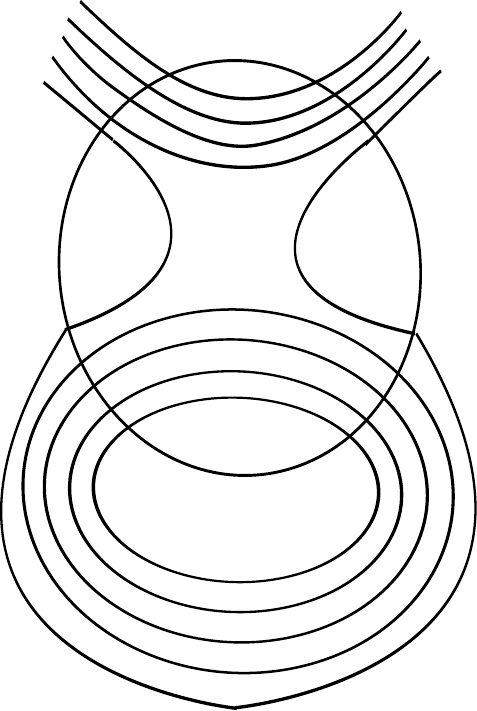}\otimes \includegraphics[angle=0, width=1.5cm, height=1.4cm]{graph96.pdf},
\eea
\bea
\Delta'(\includegraphics[angle=0, width=1.5cm, height=1.4cm]{graph104.pdf})=\includegraphics[angle=0, width=1.5cm, height=1.4cm]{graph94.pdf}\otimes \includegraphics[angle=0, width=1.5cm, height=1.4cm]{graph96.pdf}.
\eea
Putting all this together leads to
\bea\label{eq:relationc2}
\Delta'(h_2^{\includegraphics[angle=0, width=1.5cm, height=1.4cm]{simplevertex.pdf}})=(2h_1^{\includegraphics[angle=0, width=1cm, height=0.1cm]{propagatorgras.pdf}}+2h_1^{\includegraphics[angle=001, width=1.5cm, height=1.4cm]{simplevertex.pdf}})\otimes h_1^{\includegraphics[angle=0, width=1.5cm, height=1.4cm]{simplevertex.pdf}}.
\eea
As was the case for \eqref{eq:relationc1}, equation \eqref{eq:relationc2}  above illustrates identity \eqref{trei} at two-loop level.

\medskip

In addition, each of these graphs is in the image of  appropriate Hochschild one-cocycles. Indeed, one has
\bea\label{eq:52} 
B_+^{\includegraphics[angle=0, width=1.2cm, height=1cm]{oneloopc.pdf}}(\includegraphics[angle=0, width=1.5cm, height=1.4cm]{graphh3.pdf})=\frac{1}{2}(\includegraphics[angle=0, width=1.3cm, height=1.3cm]{graphh9.pdf}), \,\mbox{ i.e }
B_+^{\includegraphics[angle=0, width=1.5cm, height=1.4cm]{graph94.pdf}}(\includegraphics[angle=0, width=1.5cm, height=1.4cm]{graph96.pdf})=\frac{1}{2}(\includegraphics[angle=0, width=1.5cm, height=1.4cm]{graph97.pdf}),
\mbox{ and so on.}
\eea

Note that the combinatorial factors $\frac{1}{2}$ 
come from the 
 definition \eqref{defbplus}.  
 
 One can check that putting all these expressions together does \emph{not} give 
 the two-loop quantum corrections of the propagator. 

Nevertheless, this issue is fixed by taking into consideration
the equivalent, in the tensorial setting, 
of  the so-called planar irregular sector of Moyal quantum field theory (see \cite{TK}). In our current setting, the corresponding graphs have external legs breaking faces of different colors. 
We call them the two-color-breaking sectors.
The action of the $B_+$ operator on these graphs is given by:
\bea \label{eq:55}
B_+^{\includegraphics[angle=0, width=1.5cm, height=1.4cm]{oneloopc.pdf}}(\includegraphics[angle=0, width=1.5cm, height=1.4cm]{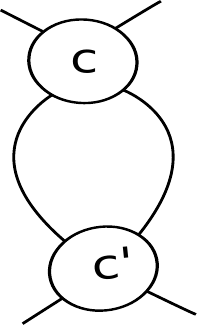})=\frac{1}{2}(\includegraphics[angle=0, width=1.5cm, height=1.4cm]{graphh4.pdf}), 
\eea
with $c\neq c'$  i.e explicitly showing strand structure
\bea \label{eq:55bis}
B_+^{\includegraphics[angle=0, width=1.5cm, height=1.4cm]{graph95.pdf}}(\includegraphics[angle=0, width=1.5cm, height=1.4cm]{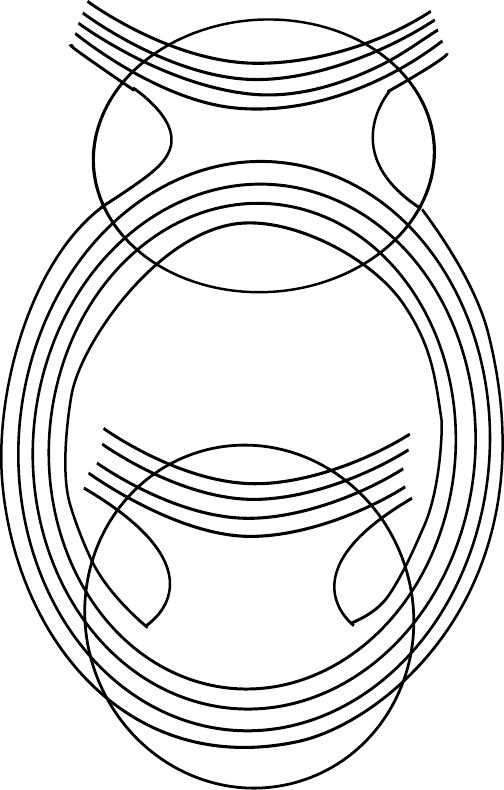})=\frac{1}{2}(\includegraphics[angle=0, width=1.5cm, height=1.4cm]{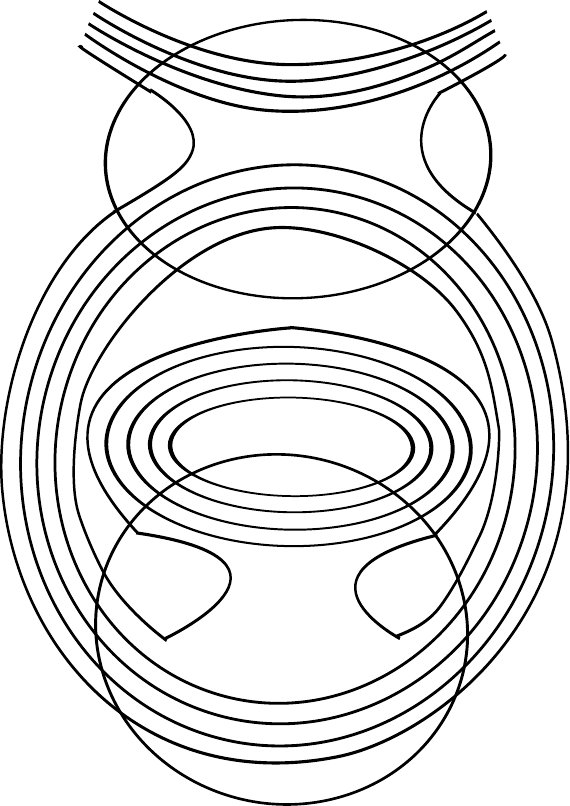}), \quad
B_+^{\includegraphics[angle=0, width=1.5cm, height=1.4cm]{graph94.pdf}}(\includegraphics[angle=0, width=1.5cm, height=1.4cm]{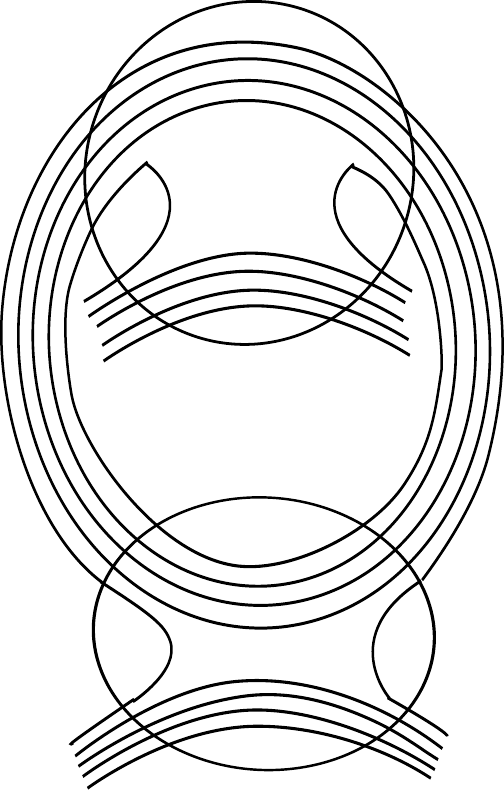})=\frac{1}{2}(\includegraphics[angle=0, width=1.5cm, height=1.4cm]{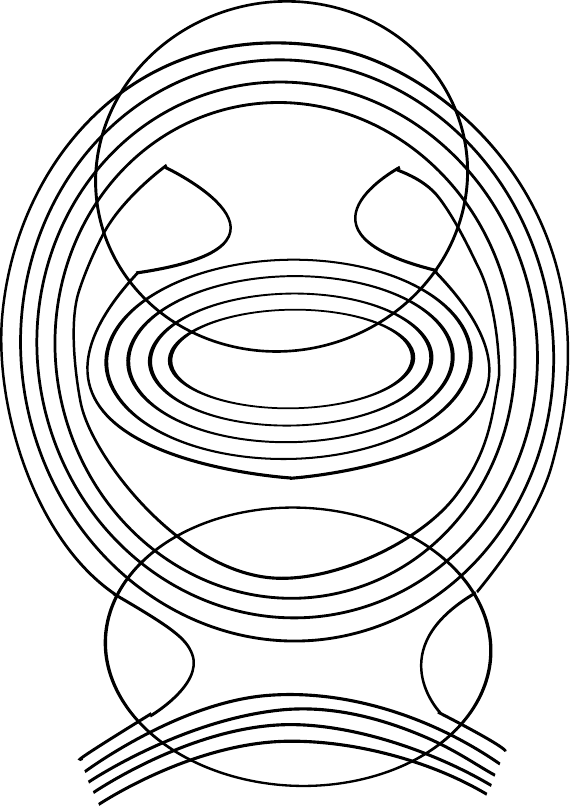}), \mbox{and so on}.
\eea
Adding the diagrams in \eqref{eq:55} to the ones in \eqref{eq:52} (with the appropriate combinatorial factors) leads to identity \eqref{unu}, at two-loop level.

\medskip

Finally, let us also check identity \eqref{doi} at two-loop level:

\bea\label{eq:theoii}
&&\Delta(B_+^{1,\includegraphics[angle=0, width=1cm, height=0.1cm]{propagatorgras.pdf}}(\sum_c \includegraphics[angle=0, width=1.5cm, height=1.4cm]{oneloopc.pdf} + \sum_c \includegraphics[angle=0, width=1.5cm, height=1.4cm]{graphh3.pdf}))\cr&=& B_+^{1,\includegraphics[angle=0, width=1cm, height=0.1cm]{propagatorgras.pdf}}(\sum_c \includegraphics[angle=0, width=1.5cm, height=1.4cm]{oneloopc.pdf} + \sum_c \includegraphics[angle=0, width=1.5cm, height=1.4cm]{graphh3.pdf})\otimes1_{\mathcal{H}}\cr&+& (id\otimes B_+^{1,\includegraphics[angle=0, width=1cm, height=0.1cm]{propagatorgras.pdf}})\Delta(\sum_c \includegraphics[angle=0, width=1.5cm, height=1.4cm]{oneloopc.pdf} + \sum_c \includegraphics[angle=0, width=1.5cm, height=1.4cm]{graphh3.pdf}).
\eea

We calculate the LHS of the relation above. One has:
\bea \label{eq:delta1}
&&\Delta(B_+^{1,\includegraphics[angle=0, width=1cm, height=0.1cm]{propagatorgras.pdf}}(\includegraphics[angle=0, width=1.5cm, height=1.4cm]{oneloopc.pdf}))=\Delta((\sum_{c'} B_+^{\includegraphics[angle=0, width=1.5cm, height=1.4cm]{graphh5.pdf}})(\includegraphics[angle=0, width=1.5cm, height=1.5cm]{oneloopc.pdf})),
\eea
and
\bea\label{eq:deltac}
\Delta((B_+^{\includegraphics[angle=0, width=1.5cm, height=1.4cm]{graphh5.pdf}})(\includegraphics[angle=0, width=1.5cm, height=1.4cm]{oneloopc.pdf}))&=&\frac{1}{2}\Delta(\includegraphics[angle=0, width=1.5cm, height=1.4cm]{graphh4.pdf})\cr&=&\frac{1}{2} \includegraphics[angle=0, width=1.5cm, height=1.4cm]{graphh4.pdf}\otimes1_{\mathcal{H}}+\frac{1}{2}1_{\mathcal{H}}\otimes\includegraphics[angle=0, width=1.5cm, height=1.4cm]{graphh4.pdf}\cr&+&\frac{1}{2}\includegraphics[angle=0, width=1.5cm, height=1.4cm]{oneloopc.pdf}\otimes \includegraphics[angle=0, width=1.5cm, height=1.4cm]{graphh5.pdf}+\frac{1}{2}\includegraphics[angle=0, width=1.5cm, height=1.4cm]{graphh3.pdf}\otimes\includegraphics[angle=0, width=1.5cm, height=1.4cm]{graphh5.pdf}.
\eea
The expression in \eqref{eq:deltac} helps to compute \eqref{eq:delta1}. Analogous expressions are computed for the other tensor Feynman graphs of equation \eqref{eq:theoii}.

In order to compute now the RHS of \eqref{eq:theoii}, we have
\bea\label{eq:Bplus1}
B_+^{1,\includegraphics[angle=0, width=1cm, height=0.1cm]{propagatorgras.pdf}}(\includegraphics[angle=0, width=1.5cm, height=1.4cm]{oneloopc.pdf})\otimes1_{\mathcal{H}}&=&(\sum_{c'}B_+^{\includegraphics[angle=0, width=1.5cm, height=1.4cm]{graphh5.pdf}})(\includegraphics[angle=0, width=1.5cm, height=1.4cm]{oneloopc.pdf})\otimes1_{\mathcal{H}}=\frac{1}{2}\sum_{c'}\includegraphics[angle=0, width=1.5cm, height=1.4cm]{graphh4.pdf}\otimes1_{\mathcal{H}}.\,\,\,\quad
\eea

One further has:
\bea\label{eq:Bplus4}
 (id\otimes B_+^{1,\includegraphics[angle=0, width=1cm, height=0.1cm]{propagatorgras.pdf}})\Delta(\includegraphics[angle=0, width=1.5cm, height=1.4cm]{oneloopc.pdf})&=& (id\otimes B_+^{1,\includegraphics[angle=0, width=1cm, height=0.1cm]{propagatorgras.pdf}})(\includegraphics[angle=0, width=1.5cm, height=1.4cm]{oneloopc.pdf}\otimes1_{\mathcal{H}} +1_{\mathcal{H}}\otimes\includegraphics[angle=0, width=1.5cm, height=1.4cm]{oneloopc.pdf})\cr&=&\includegraphics[angle=0, width=1.5cm, height=1.4cm]{oneloopc.pdf}\otimes\sum_{c'}\includegraphics[angle=0, width=1.5cm, height=1.4cm]{graphh5.pdf}+1_{\mathcal{H}}\otimes(\frac{1}{2}\sum_{c'}\includegraphics[angle=0, width=1.5cm, height=1.4cm]{graphh4.pdf}).
\eea

Summing all the contributions of type \eqref{eq:Bplus1} and  \eqref{eq:Bplus4}, we obtain the RHS of \eqref{eq:theoii}. Nevertheless, one checks that this is not equal to the LHS. 
Once again, the two-color-breaking sector saves the day.
Let us compute the LHS of \eqref{eq:theoii} for this two-breaking face sector. One has for ($c\neq c'$):
\bea\label{eq:delta4}
\Delta(B_+^{1,\includegraphics[angle=0, width=1cm, height=0.1cm]{propagatorgras.pdf}}(\includegraphics[angle=0, width=1.5cm, height=1.4cm]{graphh8.pdf}))&=&\Delta((\sum_cB_+^{\includegraphics[angle=0, width=1.5cm, height=1.4cm]{oneloopc.pdf}})(\includegraphics[angle=0, width=1.5cm, height=1.4cm]{graphh8.pdf}))=\frac{1}{2}\Delta(\includegraphics[angle=0, width=1.5cm, height=1.4cm]{graphh4.pdf})  \cr&=&\frac{1}{2}\includegraphics[angle=0, width=1.5cm, height=1.4cm]{graphh4.pdf} \otimes  1_{\mathcal{H}}+\frac{1}{2}1_{\mathcal{H}}\otimes\includegraphics[angle=0, width=1.5cm, height=1.4cm]{graphh4.pdf}\cr&+&\frac{1}{2}\includegraphics[angle=0, width=1.5cm, height=1.4cm]{oneloopc.pdf}\otimes \includegraphics[angle=0, width=1.5cm, height=1.4cm]{graphh5.pdf}.
\eea

Let us now compute the RHS of \eqref{eq:theoii} for the two-color-breaking sector. One has
\bea\label{eq:bplus3}
B_+^{1,\includegraphics[angle=0, width=1cm, height=0.1cm]{propagatorgras.pdf}}(\includegraphics[angle=0, width=1.5cm, height=1.4cm]{graphh8.pdf})\otimes1_{\mathcal{H}}&=&(\sum_cB_+^{\includegraphics[angle=0, width=1.5cm, height=1.4cm]{oneloopc.pdf}})(\includegraphics[angle=0, width=1.5cm, height=1.4cm]{graphh8.pdf})\otimes1_{\mathcal{H}}\cr &=&\frac{1}{2}\includegraphics[angle=0, width=1.5cm, height=1.4cm]{graphh4.pdf}\otimes1_{\mathcal{H}}
\eea
and
\bea\label{eq:Bplus7}
 (id\otimes B_+^{1,\includegraphics[angle=0, width=1cm, height=0.1cm]{propagatorgras.pdf}})\Delta(\includegraphics[angle=0, width=1.5cm, height=1.4cm]{graphh8.pdf})&=& (id\otimes B_+^{1,\includegraphics[angle=0, width=1cm, height=0.1cm]{propagatorgras.pdf}})(\includegraphics[angle=0, width=1.5cm, height=1.4cm]{graphh8.pdf}\otimes1_{\mathcal{H}} +1_{\mathcal{H}}\otimes\includegraphics[angle=0, width=1.5cm, height=1.4cm]{graphh8.pdf})\cr&=&1_{\mathcal{H}}\otimes(\frac{1}{2}\includegraphics[angle=0, width=1.5cm, height=1.4cm]{graphh4.pdf}).
\eea

In a similar way, one can compute the corresponding contributions of all such two-color-breaking Feynman graphs.

Adding all such contributions, identity \eqref{doi} holds!

\subsection{The case of sixth order interaction tensor model}
In this subsection, we show that the algebraic results of the previous subsection do not hold for sixth order interaction QFT models. 
We focus here on the tensor model introduced in \cite{BGR}.

In this (renormalizable) model one has both sixth order and fourth order interactions. Thus one has superficially divergent graphs such as 
$$\cG=\includegraphics[angle=0, width=1.5cm, height=1.4cm]{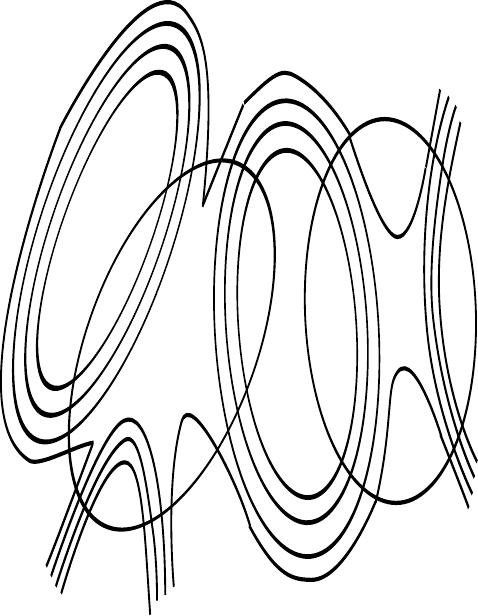}.$$
A subdivergence of this graph is given by 
$$\gamma=\includegraphics[angle=0, width=1.5cm, height=1.4cm]{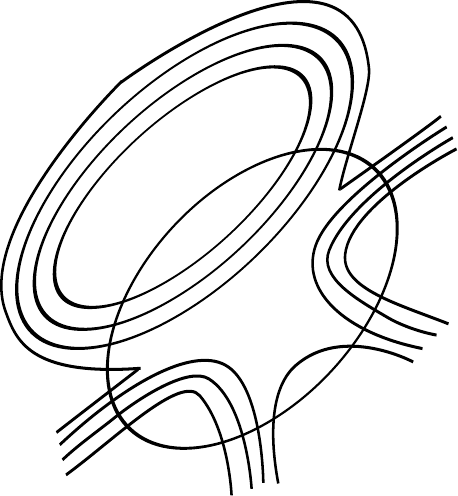}.$$ 
The contraction of $\gamma$ in $\cG$ gives
\bea
\label{problema}
\cG/\gamma = \includegraphics[angle=0, width=1.5cm, height=1.4cm]{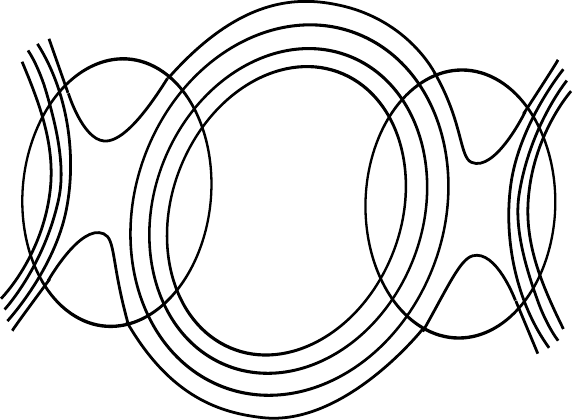}.
\eea
Nevertheless, the tensor graph $\cG/\gamma$ is not superficially divergent, in the bare theory. Since the tensor graph $\cG$ above belongs to $h_2^{\includegraphics[angle=0, width=1.5cm, height=1.4cm]{simplevertex.pdf}}$ and the graph \eqref{problema} does not belong to $h_1^{\includegraphics[angle=0, width=1.5cm, height=1.4cm]{simplevertex.pdf}}$, one can obviously not obtain identities of type \eqref{trei}. The same type of problem appears for identities \eqref{unu}
 and \eqref{doi} illustrated in the previous subsection. 
 
 Finally, let us emphasize here that this is not related in any way to the tensor character of the models discussed in this paper - the exact same behavior appears in the case of a standard $\phi^6$ theory where the one-loop correction to the $\phi^4$ term is not superficially divergent in the bare theory.

\section*{Acknowledgments}
A. Tanasa is partially funded by the grants ANR JCJC CombPhysMat2Tens and PN 09 37 01 02. R. C. Avohou kindly acknowledges the Association pour la Promotion Scientifique de l'Afrique and the Daniel Iagolnitzer Foundation, the Universit\'e Paris 13, Sorbonne Paris Cit\'e and the ANR JCJC CombPhysMat2Tens for financing his stay in Villetaneuse, stay during which this work has been done.

\appendix
\section{Proof of identity \eqref{beta3}}
In the parametric representation
\bea
\frac{1}{(p^2+m_r^2)^2}=\int_0^\infty e^{-\alpha(p^2+m_r^2)}\alpha d\alpha .
\eea
Commuting the sum and integral we obtain
\bea
I= \sum_{p\in[-N,N]^4}\frac{1}{(p^2+m_r^2)^2}=\int_0^\infty e^{-\alpha m_r^2} \alpha d\alpha 
\sum_{p\in[-N,N]^4}e^{-\alpha p^2} =\int_0^\infty \alpha d\alpha e^{-\alpha m_r^2}\Big[\sum_{p=-N}^Ne^{-\alpha p^2}\Big]^4
\eea 
Let us write
\bea 
I &=& I_1 + I_2 + I_3, \quad I_1 =\int_0^{N^{-2}} \alpha d\alpha e^{-\alpha m_r^2}\Big[\sum_{p=-N}^N  e^{-\alpha p^2}\Big]^4 \nonumber
\\
I_2 &=& \int_{N^{-2}}^1 \alpha d\alpha e^{-\alpha m_r^2}\Big[\sum_{p=-N}^Ne^{-\alpha p^2}\Big]^4, \quad I_3 =\int_1^\infty \alpha d\alpha e^{-\alpha m_r^2}\Big[\sum_{p=-N}^N  e^{-\alpha p^2}\Big]^4 .
\eea
The first term $I_1$ is bounded as $N \to \infty$. Indeed
$[\sum_{p=-N}^N e^{-\alpha p^2}\Big]^4\leq (2N+1)^4$, hence
\bea
I_1 \leq 
(2N+1)^4\int_0^{N^{-2}} \alpha d\alpha e^{-\alpha m_r^2}\leq(2N+1)^4 \frac{N^{-4}}{2}
\eea
is bounded as $N \to \infty$. 

The function $e^{-\alpha x^2}$ is decreasing and continuous on $[0,+\infty[$. Therefore 
\bea
\int_0^{N+1}e^{-\alpha t^2}dt\leq \sum_{p=0}^N e^{-\alpha p^2}\leq 1+ \int_0^Ne^{-\alpha t^2}dt, 
\eea
hence
\bea \label{goo}
-1 + \int_{-N-1}^{N+1}e^{-\alpha t^2}dt \leq \sum_{p=-N}^N e^{-\alpha p^2}\leq 1+ \int_{-N}^Ne^{-\alpha t^2}dt \le 1 +  \int_{-\infty}^{+\infty}e^{-\alpha t^2}dt = 1 + \sqrt{\frac{\pi}{\alpha}}.
\eea

The third term $I_3$ is therefore also bounded as $N \to \infty$. Indeed
\bea
I_3 \leq \int_1^\infty \alpha d\alpha e^{-\alpha m_r^2}\Big[1 +\sqrt\frac{\pi}{\alpha}\Big]^4\leq(1+\sqrt\pi)^4\int_1^\infty \alpha d\alpha e^{-\alpha m_r^2},
\eea
which is a convergent integral for $m_r^2 >0$.

We now concentrate on $I_2$ . We need  to compute exactly the coefficient of the log divergent part as $N \to \infty$.
First we have from \eqref{goo}
\bea I_2  &\le&  \int_{N^{-2}}^1   \alpha d\alpha e^{-\alpha m_r^2}  (1 + \sqrt{\frac{\pi}{\alpha}})^4  \le   \int_{N^{-2}}^1   \alpha d\alpha  (1 + \sqrt{\frac{\pi}{\alpha}})^4 \nonumber\\
&=& \pi^2 \int_{N^{-2}}^1   \frac{d\alpha}{\alpha }  (1 + \sqrt{\frac{\alpha}{\pi}})^4  = 2 \pi^2 \log N   +f_1 (N),
\eea
where $f_1 (N)$ is bounded as $N \to \infty$.
Indeed for $1 \le q \le 4$ we have that 
\bea \pi^2\int_{N^{-2}}^1   \frac{d\alpha}{\alpha } e^{-\alpha m_r^2} \sqrt{\frac{\alpha^q}{\pi^q}}  \le \pi^2\int_0^1   \frac{d\alpha}{\sqrt \alpha } . \label{goo2}
\eea 

Let us remark that 
\bea   \int_{-N-1}^{N+1}e^{-\alpha t^2}dt  =  \sqrt{\frac{\pi}{\alpha}}  [1 -  {\rm Erfc} ( \sqrt \alpha (N +1 ) ]
\eea
where ${\rm Efrc} (x) = \frac{2}{\sqrt\pi} \int_x^\infty e{-t^2}dt $ is the complementary error function. 
Then, again from \eqref{goo}
\bea I_2  &\ge&  \pi^2\int_{N^{-2}}^1   \frac{d\alpha}{\alpha } e^{-\alpha m_r^2}  [1   - \sqrt{\frac{\alpha}{\pi}}  - {\rm Erfc} [ \sqrt \alpha (N +1 ) ] )^4  . \label{goo1}
\eea
But we have, again for $1 \le q \le 4$,
\bea  \pi^2\int_{N^{-2}}^1   \frac{d\alpha}{\alpha } e^{-\alpha m_r^2}   [{\rm Erfc} ( \sqrt \alpha (N +1 ))]^q \le  \pi^2\int_{N^{-2}}^1   \frac{d\alpha}{\alpha }\label{goo3}
e^{ - \alpha  N^2} =   \pi^2\int_1^{N^2}  \frac{du}{u} e^{ - u}  \le \pi^2\int_1^\infty \frac{du}{u} e^{ - u} 
\eea
where we used the classic bound ${\rm Erfc} (x) \leq e^{-x^2}$ and changed of variables, putting $u = \alpha N^2$. This last integral is convergent as $N \to \infty$.

Therefore expanding $ [1   - \sqrt{\frac{\alpha}{\pi}}  - {\rm Erfc} [ \sqrt \alpha (N +1 ) ] )^4 $ and bounding all terms except the first in the manner of \eqref{goo2} and \eqref{goo3}
we have shown that 
\bea I_2  &\ge&  \pi^2\int_{N^{-2}}^1   \frac{d\alpha}{\alpha } e^{-\alpha m_r^2} - f_2 (N) ,
\eea
where $f_1 (N)$ is bounded as $N \to \infty$.

Finally, since $e^{-\alpha m_r^2}\geq 1-m_r^2\alpha$, 
we have
\bea
\pi^2\int_{N^{-2}}^1   \frac{d\alpha}{\alpha } e^{-\alpha m_r^2} \ge    2 \pi^2 \log N   -  f_3(N)
\eea
where $f_3 (N)$ is bounded as $N \to \infty$.
Hence $I_2 =   2 \pi^2 \log N  + f_4(N)$ with $f_4(N)$ bounded as $N \to \infty$, and $I=   2 \pi^2 \log N  + f(N)$ with $f(N)$ bounded as $N \to \infty$. This 
 completes the proof of identity \eqref{beta3}.

\begin{center}
\rule{3cm}{0.01cm}
\end{center}

\end{document}